\documentclass[twocolumn]{aastex701}
\usepackage{graphicx}
\usepackage{dcolumn}
\usepackage{bm}
\usepackage{fancyvrb}
\usepackage{breakurl}
\usepackage{multirow}
\usepackage{xcolor}
\VerbatimFootnotes
\usepackage{enumitem}
\usepackage{xspace}
\usepackage{amsmath}

\newcommand{\txs}{TXS\,0506+056\xspace}
\newcommand{\gb}{GB6\,J1542+6129\xspace}
\newcommand{\cgcg}{CGCG\,420-015\xspace}

\newcommand{\nustar}{\textit{NuSTAR}\xspace}
\newcommand{\fermi}{\textit{Fermi}\xspace}

\begin{document}

\title{Correlation Between Hard X-Ray and Cosmic Neutrino Sources: From Obscured AGN to Blazars}

\author[0000-0003-2769-3591]{Emma Kun}
\altaffiliation{Corresponding author.}
\affiliation{Department of Astronomy, Institute of Physics and Astronomy,
ELTE E\"otv\"os Lor\'and University, P\'azm\'any P\'eter s\'et\'any 1a, Budapest, Hungary}
\affiliation{Konkoly Observatory, HUN-REN Research Centre for Astronomy
and Earth Sciences, Konkoly Thege Mikl\'os \'ut 15--17, H-1121
Budapest, Hungary}
\affiliation{CSFK, MTA Centre of Excellence, Konkoly Thege Mikl\'os \'ut
15--17, H-1121 Budapest, Hungary}
\affiliation{Theoretical Physics IV, Faculty for Physics \& Astronomy,
Ruhr University Bochum, 44780 Bochum, Germany}
\email[show]{kun.emma@csfk.org}

\author[0000-0001-5607-3637]{Imre Bartos}
\affiliation{Department of Physics, University of Florida,
PO Box 118440, Gainesville, FL 32611-8440, USA}
\email{imrebartos@ufl.edu}

\author[0000-0001-5231-2645]{Claudio Ricci}
\affiliation{Department of Astronomy, University of Geneva, 1205 Geneva, Switzerland}
\affiliation{Instituto de Estudios Astrof\'isicos, Facultad de Ingenier\'ia y Ciencias, Universidad Diego Portales, Av. Ej\'ercito Libertador 441, Santiago, Chile}
\email{claudio.ricci.astro@gmail.com}

\author[0000-0002-5761-2417]{Santiago del Palacio}
\affiliation{Department of Physics and Astronomy
Chalmers University of Technology
SE-412 96 Gothenburg, Sweden}
\email{santiago.delpalacio@chalmers.se}

\author[0000-0001-6224-2417]{Francis Halzen}
\affiliation{Department of Physics, University of Wisconsin,
Madison, WI 53706, USA}
\email{francis.halzen@icecube.wisc.edu}

\author[0000-0002-1748-7367]{Julia Becker Tjus}
\affiliation{Theoretical Physics IV, Faculty for Physics \& Astronomy,
Ruhr University Bochum, 44780 Bochum, Germany}
\affiliation{Ruhr Astroparticle And Plasma Physics Center (RAPP Center),
Ruhr University Bochum, 44780 Bochum, Germany}
\affiliation{Department of Physics and Astronomy
Chalmers University of Technology
SE-412 96 Gothenburg, Sweden}
\email{julia.tjus@rub.de}

\author[0000-0003-3948-6143]{Peter L.\ Biermann}
\affiliation{Max Planck Institute for Radio Astronomy,
53121 Bonn, Germany}
\affiliation{Department of Physics \& Astronomy, University of Alabama,
Tuscaloosa, AL 35487, USA}
\email{plbiermann@mpifr-bonn.mpg.de}

\author[0000-0002-5605-2219]{Anna Franckowiak}
\affiliation{Astronomical Institute, Faculty for Physics \& Astronomy,
Ruhr University Bochum, 44780 Bochum, Germany}
\email{anna.franckowiak@rub.de}

\begin{abstract}
The origin of high-energy astrophysical neutrinos remains a key open
question in multimessenger astrophysics. A correlation between
unabsorbed hard X-ray and high-energy neutrino luminosity has been
reported in six active galactic nuclei with the highest individual
IceCube significances, linking neutrino production to compact,
photon-rich environments near supermassive black holes. We study
whether the threshold-near IceCube excesses associated with seven
\nustar-observed blazars are statistically consistent with that
relation. Calibrating the $L_\mathrm{hX}$--$L_\nu$ relation on the six
published sources via a Bayesian regression with errors on both axes,
the slope is consistent with $\beta = 1$ and the intrinsic scatter is
$\sim 0.6$\,dex. All seven new blazars are posterior-predictively
consistent with this calibration ($\chi^2_7 = 1.58$, $p = 0.98$) under
the working hypothesis that the published IceCube $\hat{n}_s$ values
reflect the signal. A null-injection test confirms that, at the present
calibration sample size, the consistency test does not by itself
adjudicate between signal and selected-background origins. A
distance-free $L_\mathrm{hX}/L_\nu$ ratio diagnostic places both
populations within the photohadronic prediction band, statistically
indistinguishable. Two diagnostics that control the common
$d_L^{\,2}$ distance bias, a redshift-partial rank correlation
($\tau|z = 0.69$, $\sim\!2.7\,\sigma$) and a flux-space permutation test
on the 13-source joint sample ($p = 6.3\times10^{-4}$, $3.23\,\sigma$),
indicate a residual $L_\mathrm{hX}$--$L_\nu$ association beyond the
distance-induced trend. We interpret these results as a conditional
consistency check; a detection-level statement requires either an
enlarged calibration set or an X-ray-weighted IceCube stacking
likelihood with internal data.
\end{abstract}


\section{Introduction}

Active galactic nuclei (AGN) are among the most powerful persistent
sources in the Universe \citep[e.g.][]{1984RvMP...56..255B,2017A&ARv..25....2P}, powered by accretion onto supermassive black
holes (SMBHs). Despite their observational diversity, AGN share a
common central engine composed of an SMBH, an accretion disk, and a
hot X-ray-emitting corona. In radio-loud systems such as blazars, this
structure is accompanied by relativistic jets aligned close to the line
of sight with superluminal speeds \citep[e.g.][]{2007Ap&SS.311..231K}, while in radio-quiet Seyfert galaxies, the emission is
dominated by the accretion flow and corona. This structural similarity
suggests that the fundamental physical processes governing energy
dissipation and particle acceleration may be common across AGN classes,
even if their observational manifestations differ.

The origin of high-energy neutrinos detected by the IceCube observatory
remains one of the central open questions in multimessenger astrophysics
\citep{2013PhRvL.111b1103A,IC2020tenyears}. Blazars have long been
considered prime candidates due to their relativistic jets and efficient
particle acceleration \citep[e.g.,][]{2022IJMPD..3130003H}. However,
the lack of a clear one-to-one association between neutrino events and
$\gamma$-ray flares \citep[e.g.][]{Kun2023}, together with the growing evidence for neutrino
emission from non-blazar AGN such as the Seyfert galaxies NGC~1068 and
NGC~4151 \citep{ngc1068_2022,Abassi2024}, suggests that the production
sites of neutrinos may not be limited to large-scale jet regions.

Recent observational and theoretical developments point toward compact
regions near the SMBH as promising sites of neutrino production. The
hot corona and inner accretion flow provide dense photon fields
favorable for photohadronic ($p\gamma$) interactions
\citep{2020PhRvL.125a1101M}; the same photon field that seeds these
interactions also renders the source opaque to the accompanying pionic
$\gamma$ rays, which are reprocessed via electromagnetic cascades into
hard X-ray emission. Generally, the X-ray emission is dominated by Comptonization in non-beamed AGN \citep[e.g.][]{1991ApJ...380L..51H}, and such cascaded emission might be on top of such baseline in their case. \citet{Neronov2024} found a possible correlation between the hard X-ray and neutrino emission for three Seyfert galaxies. \citet[hereafter K2024]{Kun2024} reported a tentative correlation
supporting the hard X-ray--neutrino correlation scenario, based on six AGN with strong individual
IceCube excesses: the Seyfert galaxies NGC~1068, NGC~4151, NGC~3079,
and \cgcg, together with the blazars \txs and \gb. Since hard X-rays
trace both coronal inverse Compton emission \citep[e.g.][]{1991ApJ...380L..51H}
and the cascade contribution from pionic $\gamma$ rays in
$\gamma$-opaque environments, this correlation is consistent with
neutrino production occurring in the corona or its immediate
environment. \citet{Arifa2025} argued that the IceCube neutrinos from TXS 0506+056 may originate not in the blazar jet but in a gamma-ray-obscured core within $\sim 10–100$ Schwarzschild radii of the central black hole, the same mechanism inferred for NGC 1068 and NGC 4151.

A correlation between millimeter and hard X-ray emission has been observed in both beamed and non-beamed AGN \citep{2022ApJ...938...87K,2023ApJ...952L..28R}, indicating that the millimeter emission probes compact regions associated with the central engine: directly the corona in radio-quiet AGN, and the corona-jet interface in radio-loud ones. The coronal origin of the compact millimeter emission in radio-quiet Seyferts is further supported by its spectral shape, which is consistent with synchrotron emission from the corona \citep{2018ApJ...869..114I}. This diagnostic is cleanest in non-beamed or misaligned systems, where the 15--55~keV emission can be cleanly associated with the corona. Our test sample, by contrast, consists of aligned blazars, whose hard X-rays can include a substantial jet inverse-Compton contribution; the calibration relation is thus established on a Seyfert-dominated, non-beamed basis and applied to an aligned-blazar test sample. We address its consequences for the emission-site interpretation and the $\kappa$ diagnostic in Sections~\ref{sec:ratio} and~\ref{sec:emissionsite}. Independent evidence linking hard X-ray
emission to neutrino production in blazars comes from a statistical
association between neutrino-detected sources and elevated hard X-ray
flux measured by SRG/ART-XC, Swift/BAT, and INTEGRAL/IBIS, suggesting
that the target photons for photohadronic interactions originate in
compact, relativistically beamed jet regions
\citep[][]{2024JCAP...05..133P}. The role of
relativistic beaming in setting the multi-messenger visibility of
such sources is dramatically illustrated by recent VLBA polarimetry
of the BL Lac PKS~1424+240, one of the highest peaks in the IceCube
9-year neutrino sky, which reveals a jet viewing angle $
0.6^\circ$, Doppler factor $\delta \sim 30$, and a coherent toroidal
magnetic-field component \citep{Kovalev2025}.

The IceCube ten-year point-source analysis
\citep{ngc1068_2022,Abassi2024} contains many candidate sources with
threshold-near excesses (local pretrial significances of order
$1$--$2\,\sigma$) whose individual reality cannot be established from
the IceCube likelihood alone. We ask whether such excesses, when
identified in blazars and combined with measured hard X-ray
luminosities, fall where the established AGN $L_\mathrm{hX}$--$L_\nu$ relation
predicts.

We organize the analysis around two distinct questions.
\textit{Question~(1): Is the underlying $L_\mathrm{hX}$--$L_\nu$ correlation
real?} This was addressed by K2024 on the six AGN with the highest
IceCube significances noted above ($R = 0.97$); we take the result as
given and use the same six sources as our calibration sample
throughout this work. \textit{Question~(2): Are the threshold-near
IceCube excesses on candidate blazars consistent with the established
relation?} We test whether seven \nustar-observed IceCube-candidate
blazars are posterior-predictively consistent with the relation
calibrated on the six K2024 sources. We are explicit that this is a
conditional consistency test, not a statement on the reality of
individual IceCube excesses against an independent background null:
the seven test sources and the six calibration sources are both
pre-selected on having $\hat{n}_s > 0$ from a common IceCube
point-source analysis, and a detection-level claim against background
fluctuations would require an X-ray-weighted IceCube stacking
likelihood with public event-level data, which is not available for
the analyzes underlying the published $\hat{n}_s$ values. We
characterize the discriminating power of the test as a function of
$\hat{n}_s$, and we calibrate it against an explicit
background-only null in Section~\ref{sec:nullcalib}.

Section~\ref{sec:xray} describes the \nustar spectral analysis, the
neutrino flux and luminosity calculation and the combined
$L_\mathrm{hX}$--$L_\nu$ sample. Section~\ref{sec:posterior} contains the
posterior-predictive test, the null calibration, and the distance-free ratio
diagnostic, which together form the central statistical analysis.
Section~\ref{sec:permutation} reports the flux-space permutation test
as a finite-sample diagnostic. Section~\ref{sec:discussion} discusses
implications and caveats; Section~\ref{sec:conclusion} concludes.

\section{Analysis and Results}
\label{sec:xray}

\subsection{\nustar Spectral Analysis}

We analyzed archival observations of the \textit{Nuclear Spectroscopic
Telescope Array} (\nustar) for a sample of seven blazars spatially
associated with IceCube neutrino source candidates having best-fit neutrino number $\hat{n}_s
> 0$ in the published ten-year point-source analyzes
\citep{ngc1068_2022,Abassi2024}. Data were reduced using standard
\texttt{NuSTARDAS} procedures, version v2.1.2 and CALDB 20230718, and
spectra from FPMA and FPMB were fitted jointly in \texttt{XSPEC}
\citep{Arnaud1996}. All sources are point-like in \nustar. Spectra were extracted within a circular aperture of 60" and background was subtracted from a source-free region.

For each observation, we adopted an absorbed power-law model of the
form
\begin{equation}
\texttt{const} \times \texttt{tbabs} \times \texttt{po},
\end{equation}
using the abundance table of \citet{wilm}. The Galactic hydrogen
column density $N_{\rm H}$ was fixed to values from the HI4PI survey
\citep{HI4PI}. A cross-normalization constant between FPMA and FPMB
was included to account for instrumental systematics, while all other
parameters were tied between the two detectors.

Unabsorbed fluxes in the $15$--$55$~keV band were derived using the
\texttt{cflux} convolution model, and luminosities were computed with
\texttt{clumin}. All uncertainties are quoted at the $1\sigma$ level.
We note that the Galactic column density is typically
$N_{\rm H} \lesssim 10^{21}$~cm$^{-2}$, so the absorption correction
is small and the distinction between observed and unabsorbed flux in
the 15--55~keV band is minor. This is in contrast to the heavily
obscured Seyfert galaxies in K2024, for which intrinsic
column densities can reach $N_{\rm H} \sim 10^{24}$~cm$^{-2}$ and the
unabsorbed flux correction is both large and model-dependent. The
blazar sample therefore provides hard X-ray luminosities that are
relatively free from the dominant absorption-correction systematic
present in the calibration sample.

All spectra are well described by a single power law, with reduced
$\chi^2$ values close to unity (Table~\ref{tab:nustar_results}). The
photon indices span $\Gamma \sim 1.4$--$2.4$, with flat-spectrum radio
quasars (FSRQs) generally exhibiting harder spectra than the BL Lac
OJ\,287, although the FSRQ range itself spans $\Gamma \sim 1.4$--$2.1$
across the sample and the IBL S2\,0109+22 has the hardest indices
($\Gamma \sim 1.5$--$1.6$). This spectral diversity within the blazar
class reflects the different positions of the SED peaks relative to
the 15--55~keV band: the FSRQs and the low-frequency-peaked BL~Lac
OJ\,287 are inverse-Compton--dominated in this band, while the
intermediate-peaked S2\,0109+22 yields the hardest indices consistent
with its IBL classification.

\begin{deluxetable*}{llllllccccc}
\tablecaption{Source properties. Source name, type, redshift ($z$),
IceCube best-fit neutrino number ($\hat{n}_s$), IceCube local pretrial
$-\log_{10} p_{\rm loc}$ (and corresponding local significance), and
\nustar spectral fit results and derived unabsorbed $15$--$55$~keV
fluxes for the seven new blazars analyzed in this work.
\label{tab:nustar_results}}
\tablehead{
\colhead{Source} &
\colhead{Type} &
\colhead{$z$} &
\colhead{$\hat{n}_s$} &
\colhead{$-\log_{10} p_{\rm loc}$} &
\colhead{ObsID} &
\colhead{Date} &
\colhead{Exposure} &
\colhead{$\Gamma$} &
\colhead{$\chi^2_\nu$} &
\colhead{$F_{15-55\,\mathrm{keV}}$} \\
&&&&&&&\colhead{(ks)} &&& \colhead{(erg~s$^{-1}$~cm$^{-2}$)}
}
\startdata
PKS 1441+25$^1$       & FSRQ & 0.940 & 3 & 0.7 (0.9$\sigma$) & 90101004002  & 2015-04-25 & 38.2  &
  $2.10\pm0.15$ & 1.13 & $(6.31\pm1.41)\times10^{-13}$ \\
3C 454.3$^2$          & FSRQ & 0.859 & 1 & 1.2 (1.6$\sigma$) & 60701034002  & 2022-06-13 & 20.8  &
  $1.46\pm0.04$ & 0.92 & $(1.49\pm0.07)\times10^{-11}$ \\
                      &      &       &   &                   & 60802038002  & 2023-05-01 & 40.2  &
  $1.48\pm0.02$ & 0.86 & $(2.67\pm0.07)\times10^{-11}$ \\
                      &      &       &   &                   & 60802038004  & 2023-05-04 & 22.4  &
  $1.45\pm0.03$ & 0.87 & $(2.62\pm0.09)\times10^{-11}$ \\
                      &      &       &   &                   & 60802038006  & 2023-05-07 & 20.7  &
  $1.44\pm0.03$ & 0.83 & $(2.58\pm0.10)\times10^{-11}$ \\
                      &      &       &   &                   & 60802038008  & 2023-05-12 & 23.0  &
  $1.47\pm0.02$ & 0.95 & $(2.66\pm0.09)\times10^{-11}$ \\
MITG J201534+3710$^3$ & FSRQ & 0.858 & 19 & 0.7 (0.9$\sigma$) & 60160732002  & 2016-05-29 & 12.7  &
  $1.54\pm0.05$ & 0.82 & $(1.17\pm0.08)\times10^{-11}$ \\
                      &      &       &    &                   & 30460021002  & 2018-11-01 & 59.6  &
  $1.65\pm0.04$ & 0.83 & $(5.02\pm0.26)\times10^{-12}$ \\
S3 0458$-$02$^4$      & FSRQ & 2.286 & 9  & 0.5 (0.4$\sigma$) & 60367003001  & 2018-04-26 & 20.7  &
  $1.73\pm0.07$ & 0.88 & $(3.91\pm0.41)\times10^{-12}$ \\
OJ 287$^5$            & BLL  & 0.306 & 16 & 0.5 (0.4$\sigma$) & 90201054002  & 2017-04-09 & 53.0  &
  $2.08\pm0.04$ & 0.95 & $(3.60\pm0.19)\times10^{-12}$ \\
                      &      &       &    &                   & 90601616002  & 2020-05-04 & 29.6  &
  $2.38\pm0.05$ & 0.91 & $(2.04\pm0.17)\times10^{-12}$ \\
                      &      &       &    &                   & 60701023002  & 2022-03-22 & 12.6  &
  $1.89\pm0.11$ & 0.75 & $(3.06\pm0.49)\times10^{-12}$ \\
                      &      &       &    &                   & 90801634002  & 2022-12-01 & 31.9  &
  $1.86\pm0.04$ & 0.96 & $(6.72\pm0.36)\times10^{-12}$ \\
                      &      &       &    &                   & 60901019002  & 2024-04-03 & 20.5  &
  $1.99\pm0.08$ & 1.01 & $(2.59\pm0.31)\times10^{-12}$ \\
S2 0109+22$^{6}$      & IBL  & 0.265 & 10 & 0.7 (0.8$\sigma$) & 60801029002  & 2022-11-22 & 57.3  &
  $1.54\pm0.05$ & 0.74 & $(3.73\pm0.28)\times10^{-12}$ \\
                      &      &       &    &                   & 60801029004  & 2022-11-24 & 100.6 &
  $1.63\pm0.04$ & 0.94 & $(3.74\pm0.19)\times10^{-12}$ \\
Ton 599$^{7}$         & FSRQ & 0.725 & 2  & 0.2 (0.0$\sigma$) & 60463037002  & 2019-05-23 & 17.0  &
  $1.97\pm0.40$ & 1.32 & $(6.11\pm4.18)\times10^{-13}$ \\
                      &      &       &    &                   & 60463037004  & 2021-06-25 & 17.6  &
  $1.59\pm0.06$ & 0.95 & $(8.21\pm0.74)\times10^{-12}$ \\
\enddata
\tablecomments{Spectra were fitted in \texttt{XSPEC} with the model
\texttt{const*tbabs*po}, adopting the abundance table of \cite{wilm}.
The Galactic hydrogen column density $N_{\rm H}$ was fixed to the HI4PI
survey value for each source \citep{HI4PI}. The multiplicative constant
accounts for cross-normalization between FPMA and FPMB. Unabsorbed
$15$--$55$~keV fluxes were derived using \texttt{cflux}; luminosities
were computed with \texttt{clumin} assuming the cosmology described in
Section~\ref{sec:nuflux}. Errors are quoted at the $1\sigma$ level.
Redshifts are from $^1$\citet{2015ApJ...815L..23A},
$^2$\citet{2022ApJS..261....2K}, $^3$\citet{2015ApJ...810...14A},
$^4$\citet{2024MNRAS.528.5990S}, $^5$\citet{1978bllo.conf..176M},
$^{6}$\citet{2016MNRAS.458.2836P}, $^{7}$\citet{2010MNRAS.405.2302H}.}
\end{deluxetable*}

To convert the observed energy flux to isotropic-equivalent luminosity,
we applied a $K$-correction appropriate for a power-law spectrum,
\begin{equation}
L_\mathrm{hX} = 4\pi d_L^2\, F_X\, (1+z)^{\Gamma_X-2},
\end{equation}
where $F_X$ is the observed 15--55~keV energy flux and $\Gamma_X$ is
the X-ray photon index from Table~\ref{tab:nustar_results}. The X-ray
log-space uncertainty $\sigma_{X,i}$ combines the measured \nustar flux
error with the $K$-correction systematic
$\sigma_{K,i} = \Delta\Gamma\cdot\log_{10}(1+z_i)$ ($\Delta\Gamma = 0.10$)
added in quadrature \citep[see][for context]{Barger2001}.

The derived unabsorbed fluxes and luminosities lie in the range
\begin{equation}
F_{15-55\,\mathrm{keV}} \sim (0.6 - 27)\times10^{-12}~
\mathrm{erg~s^{-1}~cm^{-2}},
\end{equation}
and
\begin{equation}
L_{15-55\,\mathrm{keV}} \sim 7.1\times10^{44} - 1.2\times 10^{47}~
\mathrm{erg~s^{-1}},
\end{equation}
respectively. Sources with multiple observations (e.g., 3C~454.3, OJ~287,
MITG~J201534+3710) show variability typically within a factor of
$\sim 2$--$3$. The photon indices remain consistent within
uncertainties across epochs, indicating that variability is dominated
by flux normalization rather than spectral changes. This epoch-to-epoch
spectral stability supports the use of individual observations as
independent flux measurements rather than requiring time-averaged
spectra.

\subsection{Neutrino Fluxes and Luminosities}
\label{sec:nuflux}

To calculate the neutrino luminosities, we follow K2024. We
assumed a power-law muon-neutrino plus antineutrino spectrum,
\begin{equation}
\phi_{\nu_\mu+\bar{\nu}_\mu}(E_\nu)=
\phi_0\left(\frac{E_\nu}{E_0}\right)^{-\gamma_\nu},
\end{equation}
with $E_0 = 1$~TeV. For sources with published best-fit signal-event
number $\hat{n}_s$ and spectral index $\hat{\gamma}_\nu$ from the
ten-year IceCube analysis \citep{ngc1068_2022}, we derived the
normalization $\phi_0$ from the IceCube effective area and detector
live-time by inverting the expression
\begin{equation}
n_s = \tau \sum_i A_{{\rm eff},i}
\int_{E_{l,i}}^{E_{u,i}}
\phi_{\nu_\mu+\bar{\nu}_\mu}(E_\nu)\,dE_\nu ,
\end{equation}
where the sums run over the tabulated energy bins. The published
ten-year point-source analysis covers full-detector IC86 data from
13~May 2011 to 29~May 2020 with declination-dependent effective
areas; following K2024, we adopt
the IC86 reference effective area tables together with the cumulative
ten-year exposure $\tau$, which provides a time-averaged
$A_{\rm eff} \times \tau$ that approximates the season-summed exposure
for the mid-declination sources in our sample to within $\sim 10$--$20\%$.
The 0.3--100~TeV integrated neutrino energy flux was then computed as
\begin{equation}
F_{\nu_\mu+\bar{\nu}_\mu}
=
\int_{0.3\,{\rm TeV}}^{100\,{\rm TeV}}
E_\nu\,\phi_{\nu_\mu+\bar{\nu}_\mu}(E_\nu)\,dE_\nu .
\end{equation}

To convert the observed energy flux to isotropic-equivalent luminosity,
we applied a $K$-correction similarly to what has been applied on the hard X-ray fluxes,
\begin{equation}
L_{\nu_\mu+\bar{\nu}_\mu}
=
4\pi d_L^2\,
F_{\nu_\mu+\bar{\nu}_\mu}\,
(1+z)^{\gamma_\nu-2},
\end{equation}
where $d_L$ is the luminosity distance and $z$ is the source redshift.
The exponent $(\gamma_\nu - 2)$ accounts for the energy dependence of
the $K$-correction; for the typical spectral indices in our sample
(best-fit astrophysical spectral indices $\gamma_\nu \sim 2$--$3$, \citet{ngc1068_2022,ngc1068_2022_dataset}), this correction is modest but
non-negligible at the highest redshifts. 

We emphasize that $L_\nu$ in this work is a band-limited luminosity, integrated over the fixed 0.3--100~TeV energy range from the IceCube power-law reconstruction and defined identically for the calibration and test samples. The adopted band is set primarily by consistency with the literature on which the calibration rests: the calibration-Seyfert neutrino fluxes are reported natively in 0.3--100~TeV, \citet{Neronov2024} quote 0.3--100~TeV fluxes for NGC~3079 and NGC~4151 and adopt the NGC~1068 TeV-band flux as their template, so re-deriving them over a different band would introduce an inconsistency with the published values. For the soft best-fit spectra of our sample ($\gamma_\nu \approx 2$--$3$; \citet{Neronov2024} find the three Seyfert spectra to be softer than $E^{-2}$, with most of the neutrino power emitted near the IceCube threshold of several hundred GeV), the integrated energy flux is concentrated well below 100~TeV, so $L_\nu$ is insensitive to the upper bound. A hard, $\gtrsim$PeV-peaking spectrum, as expected for pure broad-line-region (BLR) target $p\gamma$ in FSRQs \citep{2014JHEAp...3...29D}, is not what IceCube reconstructs for these sources. We quantify the effect of extending the energy range from 0.3 TeV up to 10~PeV in Appendix~\ref{app:robustness}.

The neutrino log-space uncertainty $\sigma_{\nu,i}$ requires more care,
and the asymmetric treatment between calibration and test samples is
the dominant systematic in this analysis. The published asymmetric uncertainties from K2024 are converted to a
symmetric log-space $\sigma_{\log L_\nu}$ via the geometric mean of
the upper and lower log-deviations, as required by the Gaussian-error
assumption of the linmix regression. For the seven new blazars, the corresponding likelihood profile is not publicly
available; we therefore start from the asymptotic Poisson form
$\sigma_{\nu,i}^{\rm Poisson} = 1/(\sqrt{\hat{n}_{s,i}}\,\ln 10)$
and apply an empirical scaling factor of $4.47$, derived from the mean
ratio of spectrum-derived to Poisson uncertainties on the six
calibration sources (range 3.7--5.2, standard deviation 0.59). The
constancy of this ratio across the calibration sample reflects a
$\sigma_{\rm syst} \propto 1/\sqrt{\hat{n}_s}$ scaling, consistent
with the simultaneous determination of $\hat{n}_s$ and
$\hat{\gamma}_\nu$ from the same event sample: both the Poisson term
and the spectral-index propagation term scale identically with
$\sqrt{\hat{n}_s}$, justifying a constant multiplicative form. 

We emphasize the regime in which this scaling is calibrated. The six
K2024 sources span $\hat{n}_s \in [5, 79]$, with median $\hat{n}_s
\approx 25$, where the IceCube profile likelihood is approximately
Gaussian. Three of the seven test sources lie outside
this regime: 3C~454.3 ($\hat{n}_s = 1$), Ton~599 ($\hat{n}_s = 2$), and
PKS~1441+25 ($\hat{n}_s = 3$). For these sources, the asymptotic
Poisson approximation is itself less accurate, and the asymmetry of the
underlying profile likelihood is more pronounced. We treat the
test-sample $\sigma_{\nu,i}$ values obtained from the $4.47\times$
scaling as Gaussian throughout the analysis; the impact of this choice
on the consistency test is quantified in Section~\ref{sec:postpred}
through the high-$\hat{n}_s$ vs.\ low-$\hat{n}_s$ split, and the
symmetry assumption is shown not to drive the qualitative outcome.

Throughout this work, luminosity distances were calculated assuming a
flat $\Lambda$CDM cosmology with $H_0 =
69.6$~km~s$^{-1}$~Mpc$^{-1}$, $\Omega_m = 0.286$, $\Omega_\Lambda =
0.714$, and $T_{\rm CMB,0} = 2.72548$~K, consistent with K2024.

\subsection{The Joint $L_\mathrm{hX}$--$L_\nu$ Sample}
\label{sec:correlation}

Figure~\ref{fig:postpred} shows the unabsorbed hard X-ray luminosity
versus the neutrino luminosity for the seven new blazars analyzed
here, together with the six sources from K2024. The seven
new blazars are visually consistent with the
trend established by the calibration sample and extend it toward
higher luminosities. We emphasize that this figure is descriptive: it
displays the joint sample but does not re-derive the
correlation. The K2024 relation, calibrated on the six AGN with the
highest IceCube significances, is taken as established for the rest of
this work. For descriptive purposes, the inverse-variance-weighted RMS scatter
of the joint 13-source sample around the one-to-one line
$L_\mathrm{hX} = L_\nu$ is $\sigma_{\rm scat}^{\rm w,\,obs} = 0.321$\,dex. 

\section{Posterior-Predictive Consistency Test}
\label{sec:posterior}

The central question of this paper is whether the seven new
candidate-blazar IceCube excesses, all near-threshold significances,
are consistent with being drawn from the established AGN $L_\mathrm{hX}$--$L_\nu$
relation. We address this in four steps:

\begin{enumerate}
\item[(i)] We recalibrate the relation using only the six K2024 sources.
This calibration uses the same data as K2024 but serves a different
purpose: rather than testing whether the correlation exists, we use it
to construct a posterior predictive distribution for $\log L_\nu$ at
any given $\log L_\mathrm{hX}$, which the new blazars can then be tested against.

\item[(ii)] For each new blazar, we compute the predictive
residual $\Delta_i = \log L_\nu^{\rm obs}{_i} - \log L_\nu^{\rm pred}{_i}$,
where $\log L_\nu^{\rm pred}{_i}$ is drawn from the posterior predictive
distribution at the source's measured $\log L_\mathrm{hX}$. We then ask whether
$\Delta_i$ is small compared to the predictive uncertainty
$\sigma_{{\rm pred},i}$, which combines intrinsic scatter and measurement errors; the relation-parameter covariance is not propagated in the plug-in form used here and is assessed separately in Appendix~\ref{app:robustness}. We also characterize where
the test retains discriminating power: in the high-$\hat{n}_s$ sources,
where the predictive band is set by $\sigma_{\rm int}$, the test can
meaningfully fail; in the low-$\hat{n}_s$ sources, the measurement
error $\sigma_{\nu,i}$ dominates and the test is uninformative by
construction.

\item[(iii)] We calibrate the test against an explicit background-only
null by injecting $\hat{n}_s$ values drawn from the IceCube
background-only distribution at the seven blazar positions, propagating
them through the same posterior-predictive pipeline, and asking whether
the resulting $\chi^2_{\rm null}$ distribution is statistically
distinguishable from the observed $\chi^2_7$. This addresses the
concern that a small calibration sample ($N_{\rm cal} = 6$) may render
the posterior-predictive band wide enough to accommodate signal and
selected-background hypotheses alike.

\item[(iv)] We check the distance-free $\log(L_\mathrm{hX}/L_\nu)$ ratio
diagnostic. The luminosity-luminosity plane shares a common
$d_L^{\,2}$ factor on both axes, which can artificially induce a
correlation among sources spanning a wide redshift range. Working in
$L_\mathrm{hX}/L_\nu$ cancels $d_L^{\,2}$ exactly, leaving a source-frame
quantity sensitive only to the underlying physics. If both populations
cluster around the same $\kappa$ value within the photohadronic
prediction band, the correlation is robust against this geometric
artifact.
\end{enumerate}

\subsection{Calibration of the Kun+2024 Relation}
\label{sec:calibration}

We model the hard X-ray--neutrino relation as
\begin{equation}
\log L_\nu = \alpha + \beta \log L_\mathrm{hX} + \epsilon,
\quad \epsilon \sim \mathcal{N}(0, \sigma_{\rm int}^2),
\label{eq:relation}
\end{equation}
where $\alpha$ and $\beta$ are the intercept and slope of the linear
relation, and $\epsilon$ is a Gaussian intrinsic scatter term with
variance $\sigma_{\rm int}^2$ that accounts for source-to-source
physical variation not captured by the deterministic $L_\mathrm{hX}$--$L_\nu$
relation. We treat $\log L_\mathrm{hX}$ as the predictor variable (i.e., the
independent variable in the regression), and $\log L_\nu$ as the
response. This direction reflects the physical scenario: in the
photohadronic reprocessing framework, the hard X-ray luminosity traces
the ambient photon field that drives both the $p\gamma$ interaction
rate and the $\gamma\gamma$ cascade fraction, so $L_\mathrm{hX}$ causally
influences $L_\nu$ rather than vice versa. The hard X-ray luminosity
is also the better-measured of the two quantities for our sample. We
calibrate Eq.~\ref{eq:relation} on the six K2024 sources
only, using the Bayesian linear regression of \citet{Kelly2007} with
measurement errors on both axes, as implemented in the
publicly-available \texttt{linmix} package (J.~Meyers,
\url{https://github.com/jmeyers314/linmix}). We use the spectrum-derived
$\sigma_{\nu,i}$ on $\log L_\nu$ for the calibration sample (see
Section~\ref{sec:nuflux}), four parallel chains, and a $K=2$ Gaussian
mixture prior on the latent $\log L_\mathrm{hX}$ distribution. Convergence is
established by the multivariate Gelman--Rubin statistic
$\hat{R} < 1.01$ on $(\alpha, \beta, \log\sigma_{\rm int}^2)$. The new
blazars are not used at this stage.

\begin{figure*}
\centering
\includegraphics[width=0.8\textwidth]{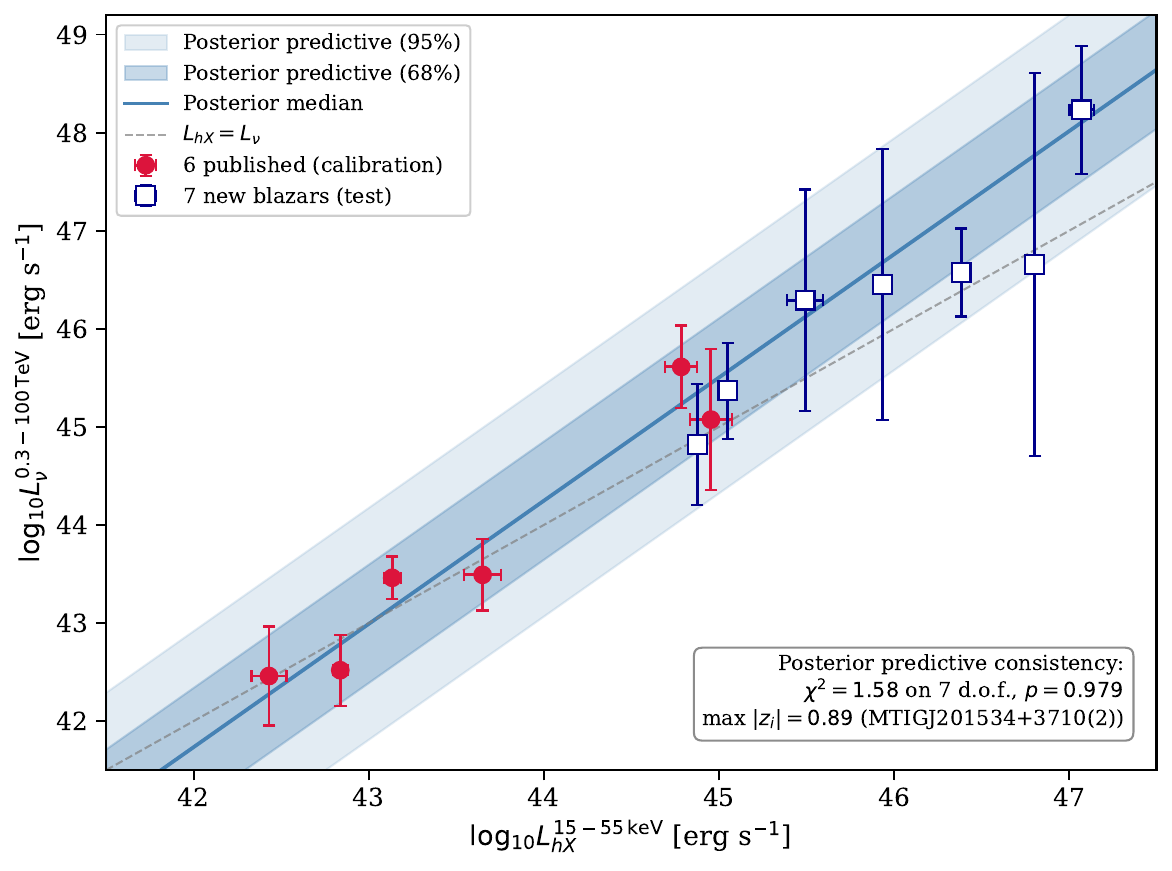}
\caption{Posterior-predictive consistency test. The blue line shows
the plug-in posterior-predictive median of the K2024
relation calibrated on the six published sources only (shown as red
filled circles), with shaded 68\% (dark) and 95\% (light) bands set by
$\pm \sigma_{\rm int,med}$ and $\pm 1.96\,\sigma_{\rm int,med}$ around
the median line. Open blue squares show the seven new blazars analyzed
in this work. The dashed line marks $L_\mathrm{hX} = L_\nu$. All seven new
sources fall within the 95\% band; per-source predictive $z$-scores
satisfy $|z_i| \leq 0.89$ (largest at MITG~J201534+3710), and the
global $\chi^2_7 = 1.58$ corresponds to $p = 0.980$. We caution that
the apparent consistency must be interpreted in light of the null
calibration of Section~\ref{sec:nullcalib}: with $\sigma_{\rm int} =
0.60$\,dex calibrated on six sources, the predictive band is wide
enough that background-only realizations also yield median
$\chi^2_{\rm null} \sim 2.5$ on this same pipeline.}
\label{fig:postpred}
\end{figure*}

\begin{table}
\centering
\footnotesize
\caption{Posterior-predictive consistency of the seven new
blazars with the K2024 calibration.
\label{tab:postpred}}
\begin{tabular}{lccccc}
\hline\hline
Source & $\hat{n}_s$ & $\log L_\mathrm{hX}^{\rm obs}$ & $\log L_\nu^{\rm obs}$ & $\log L_\nu^{\rm pred}$ & $z_i$ \\
\hline
3C 454.3              &  1 & 46.80 & 46.65 & $47.77 \pm 0.60$ & $-0.55$ \\
S2 0109+22            & 10 & 44.88 & 44.82 & $45.34 \pm 0.60$ & $-0.60$ \\
PKS 1441+25           &  3 & 45.49 & 46.29 & $46.12 \pm 0.61$ & $+0.13$ \\
MITG                  & 19 & 46.38 & 46.57 & $47.24 \pm 0.60$ & $-0.89$ \\
\,J201534+3710        &    &       &       &                  &         \\
S3 0458$-$02          &  9 & 47.07 & 48.23 & $48.11 \pm 0.60$ & $+0.14$ \\
OJ 287                & 16 & 45.05 & 45.37 & $45.57 \pm 0.60$ & $-0.26$ \\
Ton 599               &  2 & 45.93 & 46.45 & $46.68 \pm 0.60$ & $-0.15$ \\
\hline
\multicolumn{6}{r}{Full sample: $\chi^2_7 = 1.58$, $p = 0.980$} \\
\multicolumn{6}{r}{High-$\hat{n}_s$ ($\hat{n}_s \geq 9$, 4 sources): $\chi^2_4 = 1.24$, $p = 0.87$} \\
\multicolumn{6}{r}{Low-$\hat{n}_s$ ($\hat{n}_s \leq 3$, 3 sources): $\chi^2_3 = 0.34$, $p = 0.95$} \\
\hline
\end{tabular}
\vspace{0.5em}
\begin{minipage}{\columnwidth}
\footnotesize
\textbf{Note.} Predicted $\log L_\nu$ values are posterior-predictive
medians with 68\% intervals; $z_i = \Delta_i / \sigma_{{\rm pred},i}$
with $\Delta_i = \log L_{\nu,i}^{\rm obs} - \log L_{\nu,i}^{\rm pred}$,
and $\sigma_{{\rm pred},i}^2 = \sigma_{\nu,i}^2 + \sigma_{\rm int,med}^2
+ \beta_{\rm med}^2\sigma_{X,i}^2$ in the plug-in posterior-predictive
formulation (Section~\ref{sec:postpred}). The 68\% interval on
$\log L_\nu^{\rm pred}$ is dominated by $\sigma_{\rm int,med} =
0.60$\,dex, which is independent of source for the plug-in choice.
The high-$\hat{n}_s$ subsample is where the test retains
discriminating power; on the low-$\hat{n}_s$ subsample
$\sigma_{\nu,i}$ dominates and the test cannot meaningfully fail
(see Section~\ref{sec:postpred}).
\end{minipage}
\end{table}

The sampler returns posterior chains for $(\alpha, \beta,
\sigma_{\rm int})$, with median values and 68\% credible intervals
$\beta = 1.26^{+0.40}_{-0.38}$, $\sigma_{\rm int} =
0.60^{+0.78}_{-0.33}~\text{dex}$, $\alpha_{\rm pivot} =
43.51^{+0.34}_{-0.35}$, where $\alpha_{\rm pivot}$ is the relation
value at $\log L_\mathrm{hX} = 43.39$ (the median of the calibration sample).
The recovered slope is consistent with $\beta = 1$ at $\sim
0.7\,\sigma$, in agreement with the $L_\mathrm{hX} \sim L_\nu$ scaling reported
by K2024. The intrinsic scatter of $0.60$\,dex is comparable to the
calibration sample's spread along the one-to-one line and is the
dominant component of the predictive uncertainty per source for the
high-$\hat{n}_s$ test sources. We note that the credible intervals on
$\beta$ and $\sigma_{\rm int}$ are wide given $N_{\rm cal} = 6$:
$\beta$ admits values from $\sim 0.9$ to $\sim 1.7$ and $\sigma_{\rm int}$ from
$\sim 0.3$ to $\sim 1.4$\,dex at $1\,\sigma$. The use of the K2024
relation as ``established'' for the rest of this work is therefore
provisional in the literal statistical sense, and an enlarged
calibration sample remains the single most impactful improvement
identifiable for follow-up work.

\subsection{Per-Source Predictive Residuals and the
High-$\hat{n}_s$ vs.\ Low-$\hat{n}_s$ Split}
\label{sec:postpred}

Having calibrated the relation on the six K2024 sources, we now ask
whether the seven new blazars fall where the calibration predicts.
The natural quantity to compute is the posterior predictive
distribution of $\log L_\nu$ at each new blazar's measured $\log L_\mathrm{hX}$:
that is, the range of $\log L_\nu$ values the calibrated relation
predicts, accounting for the intrinsic scatter $\sigma_{\rm int}$ and the measurement error on the predictor $\log L_\mathrm{hX}$. In the plug-in form adopted below (Eq.~\ref{eq:sigmapred}) the relation parameters ($\alpha$,$\beta$,$\sigma_{\rm int}$) are fixed at their posterior medians and their covariance is not propagated; the impact of including the ($\alpha$,$\beta$) covariance is assessed in Appendix~\ref{app:robustness}.
For each new blazar $i$ we draw $M$ samples from the plug-in posterior
predictive distribution
\begin{equation}
\begin{aligned}
\log L_{\nu,i}^{\rm pred} \mid \log L_{X,i}^{\rm obs}
&= \alpha_{\rm med} + \beta_{\rm med} \,\xi_i +
\mathcal{N}(0, \sigma_{\rm int,med}^2), \\
\xi_i &\sim \mathcal{N}(\log L_{X,i}^{\rm obs}, \sigma_{X,i}^2),
\end{aligned}
\end{equation}
where $\xi_i$ marginalizes over the latent true X-ray luminosity given
the measurement error, and $(\alpha_{\rm med}, \beta_{\rm med},
\sigma_{\rm int,med})$ are fixed at their posterior medians from the
calibration chain (the plug-in choice; see below for justification). We adopt
this plug-in form because the posterior of $\sigma_{\rm int}$ at
$N_{\rm cal} = 6$ has a heavy upper tail under the Jeffreys-style prior
on $\sigma_{\rm int}^2$; full marginalization over the calibration
chain would let that tail dominate $\sigma_{{\rm pred},i}$ and render
the test uninformative, whereas the plug-in choice retains the
representative scatter recovered by the regression
\citep[cf.][]{Hogg2010}. The predictive residual is
\begin{equation}
\Delta_i = \log L_{\nu,i}^{\rm obs} - \log L_{\nu,i}^{\rm pred},
\end{equation}
and the full predictive variance per source combines the observed
neutrino measurement error and the spread of the predictive
distribution,
\begin{equation}
\sigma_{{\rm pred},i}^2
= \sigma_{\nu,i}^2
+ \sigma_{\rm int,med}^2
+ \beta_{\rm med}^2 \sigma_{X,i}^2.
\label{eq:sigmapred}
\end{equation}
We summarize consistency through the predictive $z$-score
$z_i = \Delta_i / \sigma_{{\rm pred},i}$ and the two-sided $p$-value
from the standard normal. A global consistency test is given by the
joint statistic
\begin{equation}
\chi^2 = \sum_{i=1}^{N_{\rm new}} z_i^2,
\end{equation}
which, under the null hypothesis that the new blazars are drawn from
the calibrated relation, follows $\chi^2_{N_{\rm new}}$.

The per-source results are given in Table~\ref{tab:postpred} and
visualized in Figure~\ref{fig:postpred}. All seven new blazars satisfy
$|z_i| \leq 0.89$, the global $\chi^2_7 = 1.58$ corresponds to
$p = 0.980$, and every source falls within the 95\%
posterior-predictive band.

The discriminating power of the test, however, is unevenly distributed
across the sample, and inspection of the dominant term in
Eq.~\ref{eq:sigmapred} makes this explicit. Eq.~\ref{eq:sigmapred} fixes $(\alpha,\beta)$ at their medians and therefore omits the relation-parameter covariance, which at the high-luminosity end is significant given the $\sim 1$--$3$\,dex extrapolation from the calibration pivot. We show in Appendix~\ref{app:robustness} that a robust treatment of the $(\alpha,\beta)$ covariance broadens $\sigma_{{\rm pred},i}$ at the high-luminosity sources (from $\sim 0.7$ to $1.4$--$2.4$\,dex) without altering the consistency conclusion, while full marginalization over the heavy-tailed posterior is uninformative. With $\sigma_{\rm int,med}
= 0.60$\,dex and the $4.47\times$ scaled Poisson form of
$\sigma_{\nu,i}$, the four sources with $\hat{n}_s \geq 9$
(MITG~J201534+3710 with $\hat{n}_s = 19$, OJ~287 with $\hat{n}_s =
16$, S2~0109+22 with $\hat{n}_s = 10$, S3~0458--02 with $\hat{n}_s =
9$) have $\sigma_{\nu,i} \in [0.45, 0.65]$\,dex, comparable to or
smaller than $\sigma_{\rm int}$, so the predictive variance is
$\sigma_{\rm int}$-dominated and a deviation from the calibrated
relation would in principle be visible. The three sources with
$\hat{n}_s \leq 3$ (3C~454.3, Ton~599, PKS~1441+25) instead have
$\sigma_{\nu,i} \in [1.1, 2.0]$\,dex, two to three times larger than
$\sigma_{\rm int}$, so their predictive variance is dominated by
$\sigma_{\nu,i}$ and any luminosity assignment compatible with
$\hat{n}_s > 0$ trivially passes the consistency test. Restricted to
the high-$\hat{n}_s$ subsample, the test gives $\chi^2_4 = 1.24$
($p = 0.87$, $\max|z_i| = 0.89$); restricted to the low-$\hat{n}_s$
subsample, $\chi^2_3 = 0.34$ ($p = 0.95$). The full $\chi^2_7 = 1.58$
should therefore be read primarily as the $\chi^2_4$ contribution of
the four high-$\hat{n}_s$ sources, with the three low-$\hat{n}_s$
sources contributing $\sigma_{\nu,i}$-dominated noise rather than
discriminating information.

We emphasize that the relevant quantity for the consistency test is $\hat{n}_s$ (through $\sigma_{\nu,i}\propto 1/\sqrt{\hat{n}_s}$), not the local IceCube significance, and the two are not monotonically related: the local test statistic depends on the fitted spectral index, the declination-dependent atmospheric background, and the energies of the contributing events, so a few hard-spectrum, high-energy events in a low-background regime can yield a higher local significance than many soft-spectrum, threshold-near events. This is why, for example, 3C~454.3 ($\hat{n}_s = 1$) reaches a higher local significance ($1.6\,\sigma$) than MITG~J201534+3710 ($\hat{n}_s = 19$, $0.9\,\sigma$; Table~\ref{tab:nustar_results}). The discriminating power of our test, set by $\sigma_{\nu,i}$, therefore resides in the high-$\hat{n}_s$ sources even though these carry the lowest local significances; there is no contradiction, since the local significance measures detection against background while $\sigma_{\nu,i}$ measures the precision of the inferred neutrino luminosity.

A separate consequence of this structure is that the $\hat{n}_s > 0$
selection cut applied to the test sample acts as a survival
filter on precisely the quantity being tested. Sources whose
underlying $\hat{n}_s$ fluctuated below zero under noise are excluded
by construction; sources at $\hat{n}_s = 1$ or $2$ pass the cut with
$\sigma_{\nu,i}$ large enough that any $L_\nu$ within
$\sim 1$--$2$\,dex of the prediction looks acceptable. Both effects
push the test toward apparent consistency independent of the truth of
the signal hypothesis. The null calibration of
Section~\ref{sec:nullcalib} quantifies this directly.

\subsection{Null-Calibration of the Consistency Test}
\label{sec:nullcalib}
\begin{figure*}
\centering
\includegraphics[width=\textwidth]{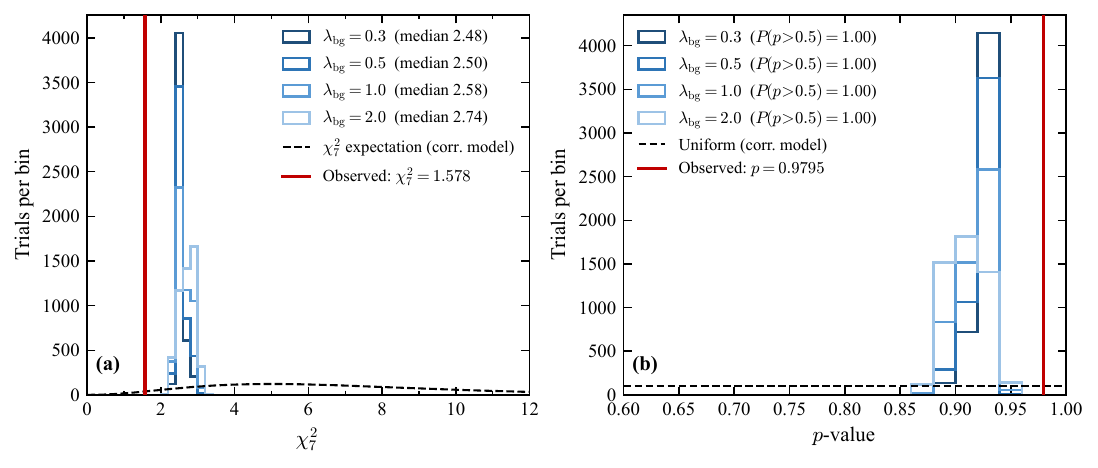}
\caption{Null calibration of the posterior-predictive consistency
test. \emph{Left:} Distribution of $\chi^2_7$ under a background-only
null in which $\hat{n}_s$ values are drawn from
${\rm Poisson}(\lambda_{\rm bg})$ truncated to $\hat{n}_s > 0$, for
four background expectations $\lambda_{\rm bg} \in \{0.3, 0.5, 1.0,
2.0\}$. The black dashed curve is the $\chi^2_7$ distribution that
would obtain if the test were correctly calibrated against the K2024
relation (median $\sim 7$). The observed value $\chi^2_7 = 1.58$ (red
vertical line) sits below the null medians ($\sim 2.5$--$2.7$). Both
the null and the observed value lie far below the correctly-calibrated
expectation, reflecting the wide predictive band that follows from
$\sigma_{\rm int} = 0.60$\,dex at $N_{\rm cal} = 6$.
\emph{Right:} Distribution of $p$-values under the same null. Median
$p_{\rm null}$ exceeds $0.9$ across all four $\lambda_{\rm bg}$;
$P(p_{\rm null} > 0.5) = 1.00$ in every case. The observed $p = 0.980$
sits in the upper tail of all four null distributions, but the
predictive band is too wide for this to constitute formal rejection
of the background hypothesis.}
\label{fig:null}
\end{figure*}
To assess whether $\chi^2_7 = 1.58$ ($p = 0.980$) is itself
informative about the reality of the seven IceCube excesses, we
calibrate the test against an explicit background-only null. The
construction is the following. We hold the X-ray luminosities
$\{\log L_{X,i}\}$ and redshifts $\{z_i\}$ of the seven test sources
fixed at their observed values, preserving the $L_\mathrm{hX}$ and $z$
distributions exactly. We then generate $\hat{n}_s$ realizations
under an IceCube background-only hypothesis,
$\hat{n}_{s,i} \sim {\rm Poisson}(\lambda_{\rm bg})$, and apply the
$\hat{n}_s > 0$ truncation that mirrors the post-hoc selection cut
imposed on the real test sample. Each null realization is propagated
through the same flux pipeline (Section~\ref{sec:nuflux}), the same
$4.47\times$ scaled Poisson $\sigma_{\nu,i}$, and the same plug-in
posterior-predictive consistency test against the calibrated K2024
relation, yielding a null distribution of $\chi^2_7$ values. We
explore $\lambda_{\rm bg} \in \{0.3, 0.5, 1.0, 2.0\}$, spanning the
range of empty-sky background expectations for declination-dependent
IceCube point-source analyzes; $5000$ realizations are drawn at each
$\lambda_{\rm bg}$. Use of the same $\hat{\gamma}_\nu$ values as the
observed sample is conservative for the null, since true background
events would have a more steeply falling spectrum.

The results are summarized in Figure~\ref{fig:null} and
Table~\ref{tab:null}. Median null $\chi^2_7$ values lie in the range
$2.48$--$2.74$ across the four $\lambda_{\rm bg}$, corresponding to
median $p_{\rm null} \in [0.91, 0.93]$. The probability that a
background-only realization produces $p_{\rm null} > 0.5$ is unity in
every case, and $P(p_{\rm null} > 0.9)$ ranges from $0.67$ to $0.97$.
Both the null and the observed distributions sit far below the
$\chi^2_7$ distribution that would obtain under a correctly-calibrated
test of the K2024 relation (median $\chi^2_7 \approx 7$, dashed line
in Figure~\ref{fig:null}, left). This reflects the fundamental
property of the predictive band: with $\sigma_{\rm int} = 0.60$\,dex
calibrated on six sources and propagated as a plug-in into
$\sigma_{{\rm pred},i}$, the band is wide enough that no realistic
seven-source sample, signal or background, populates the upper tail
of the $\chi^2_7$ expectation. The test is conservative in both
directions: it does not falsely reject the calibrated relation when
the relation holds, but it also does not reject background.

\begin{table}
\centering
\footnotesize
\caption{Null calibration of $\chi^2_7$ under background-only
$\hat{n}_s$ realizations.
\label{tab:null}}
\begin{tabular}{lcccc}
\hline\hline
$\lambda_{\rm bg}$ & med. $\chi^2_{\rm null}$ & med. $p_{\rm null}$
& $P(p_{\rm null} > 0.5)$ & $P(p_{\rm null} > 0.9)$ \\
\hline
0.3 & 2.48 & 0.93 & 1.00 & 0.97 \\
0.5 & 2.50 & 0.93 & 1.00 & 0.94 \\
1.0 & 2.58 & 0.92 & 1.00 & 0.83 \\
2.0 & 2.74 & 0.91 & 1.00 & 0.67 \\
\hline
Obs. & 1.58 & 0.98 & --- & --- \\
\hline
\end{tabular}
\vspace{0.5em}
\begin{minipage}{\columnwidth}
\footnotesize
\textbf{Note.} $5000$ background-only realizations per
$\lambda_{\rm bg}$. The observed $\chi^2_7 = 1.58$ sits below the
null median for all four values, and $P(\chi^2_{\rm null} \leq
\chi^2_{\rm obs}) < 10^{-3}$ in every case. The same predictive band
that prevents the test from rejecting the calibrated relation also
prevents it from rejecting background; both lie in a regime where the
predictive variance dominates the per-source residual. A meaningful
rejection requires an enlarged calibration sample (which would
tighten $\sigma_{\rm int}$) or an alternative test statistic that is
less band-dominated.
\end{minipage}
\end{table}

We draw two conclusions from this calibration. First, the formal
$\chi^2_7$ statistic does not, on its own, distinguish between the
hypothesis that the seven IceCube excesses reflect signal and the
hypothesis that they are selected background fluctuations: both
populate the same low-$\chi^2$, high-$p$ region of the test
distribution. Second, the observed value $\chi^2_7 = 1.58$ is
nonetheless lower than the null median by approximately one unit, a
direction consistent with signal but not statistically distinguishable
from background within the precision afforded by the present
calibration. The test as constructed is therefore best read as a conditional consistency check: under the assumption that the
IceCube excesses reflect signal, the seven blazars are statistically
indistinguishable from the K2024 calibration; under the alternative
that they are selected background fluctuations, the test does not
reject that hypothesis either. Discriminating between these two
readings requires either a tighter $\sigma_{\rm int}$ from an enlarged
calibration sample, or an independent test of detection-level
significance via X-ray-weighted IceCube stacking with public
event-level data.

\subsection{Distance-Free $L_\mathrm{hX}/L_\nu$ Ratio Diagnostic}
\label{sec:ratio}
\begin{figure*}
\centering
\includegraphics[width=0.7\textwidth]{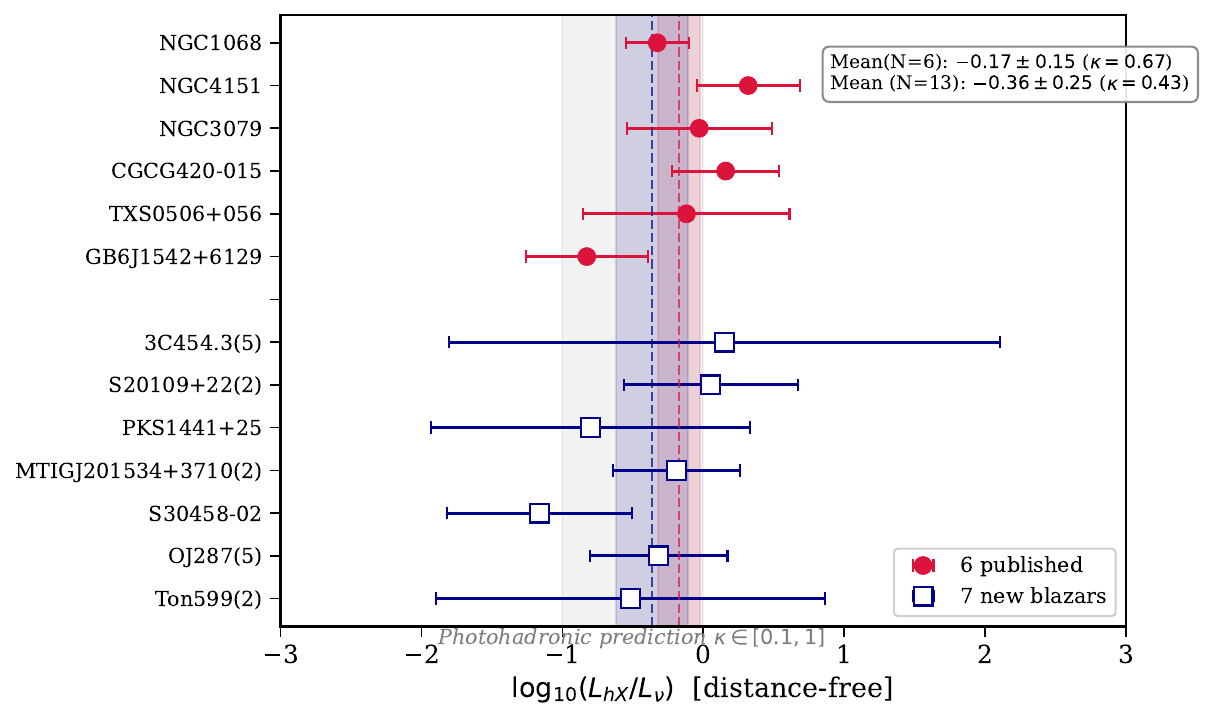}
\caption{Distance-free $\log(L_\mathrm{hX}/L_\nu)$ ratio diagnostic. Red filled
circles: six K2024 calibration sources; open blue squares:
seven new blazars from this work. Vertical dashed lines show
inverse-variance-weighted means with shaded 68\% bands. The grey band
indicates the photohadronic prediction range $\kappa \in [0.1,1]$.
Population-weighted $\kappa$ values are $0.67$ (calibration) and
$0.43$ (new sample); the two means differ by $-0.19 \pm 0.29$,
consistent with zero at $\sim 0.7\,\sigma$. Error bars per source
reflect the spectrum-derived $\sigma_{\nu,i}$ for the calibration
sample and the empirically scaled Poisson approximation for the test
sample (Section~\ref{sec:nuflux}). The agreement is necessary but not
sufficient evidence of common physics, since the joint \nustar$+$IceCube
selection envelope preferentially populates the same $\kappa$ window
that the photohadronic prediction occupies; see
Section~\ref{sec:ratio}.}
\label{fig:ratio}
\end{figure*}
A second, independent diagnostic uses the source-frame ratio
\begin{equation}
\log\!\left(\frac{L_\mathrm{hX}}{L_\nu}\right)
= \log\!\left(\frac{F_X}{F_\nu}\right)
+ (\Gamma_X - \gamma_\nu)\log_{10}(1+z),
\end{equation}
which cancels the common $d_L^{\,2}$ factor exactly. The
photohadronic reprocessing framework predicts $\kappa = L_\mathrm{hX}/L_\nu \in
[0.1, 1]$, set by the cascade development and photon escape fraction
\citep{2020PhRvL.125a1101M, 2020ApJ...891L..33I, 2022ApJ...939...43E};
the relevant physics lives entirely in source-frame quantities and is
unaffected by sample distance distribution. Figure~\ref{fig:ratio}
shows the per-source ratios for the calibration and test samples.
Inverse-variance-weighted means are
\begin{align}
\langle\log(L_\mathrm{hX}/L_\nu)\rangle_{\rm pub} &=
-0.17 \pm 0.15 \quad (\kappa \approx 0.67),\\
\langle\log(L_\mathrm{hX}/L_\nu)\rangle_{\rm new} &=
-0.36 \pm 0.25 \quad (\kappa \approx 0.43).
\end{align}
Both population-weighted means lie within the photohadronic
prediction band $\kappa \in [0.1,1]$, and the two means differ by
$-0.19 \pm 0.29$, consistent with zero at $\sim 0.7\,\sigma$. The
seven new blazars are therefore statistically indistinguishable from
the published sample in their distance-free X-ray-to-neutrino
luminosity ratio.

We caution that, like the $\chi^2$ test, the ratio diagnostic is
constructed on samples conditioned on simultaneous \nustar detection
and IceCube $\hat{n}_s > 0$. This joint sensitivity envelope
preferentially populates the very $\kappa$ range that the
photohadronic prediction occupies: a source with $\kappa$ outside
$[0.1, 1]$ at typical distances would fall below one of the two
detection thresholds and be absent from any sample so constructed.
Population-averaged agreement at $0.7\,\sigma$ within this band is
therefore necessary but not sufficient evidence of common physics.
The diagnostic places upper limits on outliers, no source in
either sample sits more than $\sim 1$\,dex from the band's edges, but it cannot, by construction, distinguish a coronal/cascade origin
from any other emission scenario whose $\kappa$ predictions overlap
the same window. Equivalently, the comparison of $\langle\kappa\rangle$ between the two samples is physically meaningful only if the hard X-rays are produced by the same mechanism in both (Section~\ref{sec:emissionsite}): a jet inverse-Compton contribution to the blazar 15--55~keV band that is unaccompanied by neutrino production would bias $\kappa_{\rm new}$ without reflecting a difference in the underlying photohadronic physics.

\section{Flux-Space Permutation Test as a Finite-Sample Diagnostic}
\label{sec:permutation}

\begin{figure*}
\centering
\includegraphics[width=\textwidth]{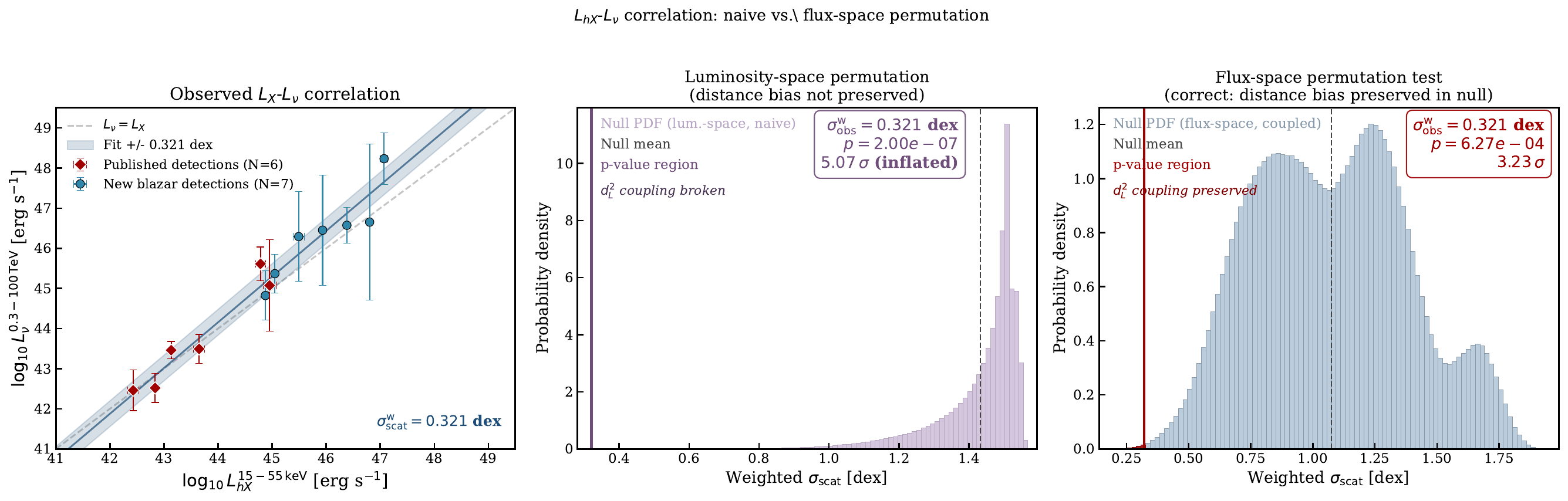}
\caption{Flux-space permutation test as a finite-sample diagnostic.
\emph{Left:} Observed joint sample with the weighted forward
regression and its $\pm\sigma_{\rm scat}^{\rm w}$ band.
\emph{Middle:} Null distribution from a luminosity-space permutation
that breaks the $d_L^{\,2}$ coupling in the null but preserves it in
the observation, yielding an inflated nominal significance of
$5.07\,\sigma$.
\emph{Right:} Null distribution from the flux-space permutation
adopted here, in which $(F_\nu, \gamma_\nu, \sigma_\nu)$ are shuffled
jointly and $L_\nu$ is recomputed at each receiving source's
redshift; this preserves the $d_L^{\,2}$ coupling identically in
observation and null, and gives the physically meaningful
$3.23\,\sigma$ ($p = 6.3\times10^{-4}$).}
\label{fig:permutation}
\end{figure*}
As a finite-sample diagnostic of the combined 13-source structure, we
perform a flux-space permutation test in which the neutrino quantities
$(F_\nu, \gamma_\nu, \sigma_\nu)$ are shuffled jointly across the
13 sources of the joint sample, breaking the observed pairing between
each X-ray source and its measured neutrino quantities while preserving
the $L_\mathrm{hX}$--$L_\nu$ pair structure under the random reassignment.
For each permutation, $\log L_\nu$ is recomputed at the receiving
source's redshift,
\begin{equation}
\label{eq:Lnuperm}
\begin{aligned}
\log L_\nu^{\rm perm}[i] &= \log F_\nu^{\rm perm}[i] \\
                        &+ \log\!\left(4\pi d_{L,i}^{\,2}\right) \\
                        &+ \bigl(\gamma_\nu^{\rm perm}[i] - 2\bigr)\,\log(1+z_i),
\end{aligned}
\end{equation}
where $F_\nu^{\rm perm}[i]$ and $\gamma_\nu^{\rm perm}[i]$ are the
neutrino flux and spectral index drawn from a random reassignment of
the 13 neutrino measurements to the 13 X-ray sources, and $d_{L,i}$,
$z_i$ are the luminosity distance and redshift of source $i$ (its
original X-ray sky position). This coupled shuffle preserves the
$d_L^{\,2}$ factor identically in observation and null, so the test asks whether the observed $L_\mathrm{hX}$ and
$L_\nu$ values are paired more tightly than random reassignment within
the same IceCube-positive sample. The test statistic is the
inverse-variance-weighted RMS scatter of $\log L_\nu$ around the
weighted forward regression of $L_\nu$ on $L_\mathrm{hX}$, with weights
$w_i = 1/(\sigma_{\nu,i}^{\,2} + \sigma_{X,i}^{\,2})$ recomputed per
permutation from the shuffled $\sigma_\nu$.
We adopt $L_\mathrm{hX}$ as the regressor because the X-ray luminosities are
substantially better constrained than the Poisson-dominated neutrino
luminosities, so this choice minimizes the regression-dilution bias
that would otherwise be incurred by using the noisier variable as the
predictor.
We use $N_{\rm perm} = 10^7$ realizations.

The observed weighted scatter is $\sigma_{\rm scat}^{\rm w,obs} = 0.321$\,dex,
compared to a null mean $\langle\sigma_{\rm scat}^{\rm w}\rangle_{\rm null} \simeq 1.08$\,dex.
The $p$-value is $p = 6.3 \times 10^{-4}$ ($3.23\,\sigma$). For comparison, a naive
luminosity-space permutation that scrambles $\log L_\nu$ directly leaves the
distance inflation in the observation but breaks it in the null, yielding
$\langle\sigma_{\rm scat}^{\rm w}\rangle_{\rm null,\,lum} \simeq 1.43$\,dex,
$p = 2.0 \times 10^{-7}$, and an artificially inflated significance of
$5.07\,\sigma$. The gap between the two tests, $\sim 1.8\,\sigma$, quantifies
the $d_L^{\,2}$ contribution that the flux-space construction removes.

The flux-space null distribution is non-Gaussian, with three visible modes
that reflect the trimodal $\log d_L^{\,2}$ distribution of the joint 13-source
sample: four nearby Seyferts at $z \lesssim 0.03$, three mid-redshift
blazars at $z \sim 0.3$, and six high-redshift blazars at $z \gtrsim 0.6$.
Under a random reassignment, the high-weight neutrino fluxes are shuffled
into one of these three distance clusters, producing three regimes of
weighted scatter. The $p$-value is therefore evaluated directly from the
empirical CDF of the permutation distribution and does not assume any
parametric form for the null.

We treat this permutation result as a secondary, finite-sample
diagnostic rather than a detection statistic, in keeping with the
framing established in the introduction. The test is conditional on
the IceCube-positive sample (it does not establish detection against
the blazar population at large), and the flux-space dynamic range
available for shuffling is limited by the clustering of the seven new
$F_\nu$ values near the IceCube threshold. Both effects are addressed
in Section~\ref{sec:caveats}. We also note that this test, unlike the
$\chi^2$ test of Section~\ref{sec:posterior}, is sensitive to the relative ordering of $L_\mathrm{hX}$ and $L_\nu$ within the joint sample
rather than to the absolute predictive band, and the two diagnostics
therefore probe partially independent aspects of the structure.
Figure~\ref{fig:permutation} shows the observed correlation alongside
both null distributions.

As an independent, rank-based check of the distance-controlled structure, we computed the partial rank correlation between $\log L_\mathrm{hX}$ and $\log L_\nu$ on the joint 13-source sample, controlling for redshift. For rank statistics this is equivalent to controlling for the common $4\pi d_L^2$ factor, since $z$ and $d_L$ are monotonically related. The partial Kendall $\tau$ \citep{Padovani1992,AkritasSiebert1996} is $\tau|z = 0.69$ (raw $\tau = 0.90$) and the partial Spearman is $\rho|z = 0.71$ (raw $\rho = 0.96$); a residual (Freedman--Lane) permutation test gives $p = 0.0066$ ($\sim 2.7\,\sigma$) for the partial Spearman, consistent with the $3.23\,\sigma$ of the flux-space test above. The reduction from the raw to the partial coefficients confirms that a substantial part of the luminosity-space correlation is distance-induced, while the surviving partial correlation reflects a real residual association. A Pearson partial controlling for $z$ linearly ($0.96$) does not remove this confound, $\log(4\pi d_L^2)$ is strongly nonlinear in $z$ across $z = 0.003$--$2.29$, whereas controlling for $\log d_L^2$ directly gives $0.85$; the rank statistics, invariant under the monotonic $z$--$d_L$ mapping, control the distance factor properly.

\section{Discussion}
\label{sec:discussion}

\subsection{Physical Interpretation}

A simple physical argument motivates the hard X-ray--neutrino
correlation within a photohadronic framework. In the
$\Delta$-resonance approximation, the threshold condition $E'_p
E'_\gamma \simeq 0.3~{\rm GeV}^2$ combined with the inelasticity
$E'_\nu \simeq E'_p / 20$ implies $E'_\gamma E'_\nu \simeq
0.015~{\rm GeV}^2$ in the source frame. Transforming to the observer
frame via $E^{\rm obs} = \delta\,E' / (1+z)$ for both photon and
neutrino, the target photon energy required for neutrino production
at $E_\nu^{\rm obs}$ becomes
\begin{equation}
E_\gamma^{\rm obs} \simeq 1.9\,{\rm keV}
\left(\frac{\delta}{10}\right)^2
\left(\frac{2}{1+z}\right)^2
\left(\frac{E_\nu^{\rm obs}}{200\,{\rm TeV}}\right)^{-1},
\label{eq:egamma_obs}
\end{equation}
where the $(2/(1+z))^2$ form is chosen so that the prefactor
corresponds to the reference values $\delta = 10$, $z = 1$, and
$E_\nu^{\rm obs} = 200$~TeV; we caution that the coefficient depends
sensitively on the adopted resonance constant and inelasticity factor
(values quoted in the literature span $\sim 1$--$10$\,keV at the same
reference parameters). Equation~\ref{eq:egamma_obs} demonstrates that
for typical neutrino-emitting blazars ($\delta \gtrsim 10$, $z \sim
0.5$--$2$), the relevant target photons lie in the soft-to-hard
X-ray band covered by \nustar at the higher-end values of $\delta$,
and shift toward the soft X-ray band for moderate Doppler factors.
The 15--55~keV \nustar band probes the high-$\delta$, high-$E_\nu$
corner of the parameter space, with a soft tail that overlaps the
\textit{Swift}/XRT and \textit{NuSTAR} sensitivity curves.

We stress that Equation~\ref{eq:egamma_obs} illustrates only the X-ray-as-target branch, the kinematics under which soft-to-hard X-ray photons can serve as the $p\gamma$ target for high-Doppler-factor sources, and is complementary to, not the basis of, the $L_\mathrm{hX}$--$L_\nu$ relation. The relation itself rests on the calorimetric reprocessing interpretation (Section~\ref{sec:emissionsite}), in which the hard X-rays are the cascade product, $L_\mathrm{hX}\approx\kappa L_\nu$, independent of the target photon field. The target field is class-dependent: the coronal X-ray field in the Seyferts, and the BLR (UV) field in the FSRQs, for which the neutrino spectrum is expected to peak at $\gtrsim$PeV \citep{2014JHEAp...3...29D}. Equation~\ref{eq:egamma_obs} and Section~\ref{sec:emissionsite} are thus mutually consistent: the hard X-rays trace the cascade reprocessing regardless of which photon field provides the $p\gamma$ target.

For characteristic compact blazar emission regions
($R' \sim 10^{15}$--$10^{16}$~cm, $\delta \sim 10$, $L_\mathrm{hX} \sim
10^{44}$--$10^{46}$~erg~s$^{-1}$), the photohadronic interaction
efficiency is $f_{p\gamma} \sim 10^{-3}$--$10^{-1}$. The maximum
particle energy attainable in such compact regions is set by the
magnetic-field energy product $Br$, which empirically saturates near
$\sim 10^{16}$\,G\,cm across nearly ten orders of magnitude in
black-hole mass for spin-down-powered systems and may be exceeded in
accreting configurations \citep[see also][]{2024APh...16102976A}. The
accompanying $\pi^0$-decay photons experience high optical depths to
pair production ($\tau_{\gamma\gamma}/f_{p\gamma} \sim 10^2$--$10^3$),
so even moderately efficient sources are opaque to their own pionic
$\gamma$-ray emission \citep[e.g.][]{2022IJMPD..3130003H,
2021ApJ...911L..18K}. The resulting electromagnetic cascades
redistribute energy toward lower frequencies, leading to the
quasi-calorimetric relation $L_\mathrm{hX} \sim \kappa\,L_\nu$, with
$\kappa \sim 0.1$--1 set by the cascade development and photon escape
fraction \citep{2020PhRvL.125a1101M, 2020ApJ...891L..33I,
2022ApJ...939...43E}. Both our calibration ($\kappa \approx 0.67$) and
test ($\kappa \approx 0.43$) populations sit within this predicted
band (Section~\ref{sec:ratio}). We emphasize, however, that the
$\kappa \in [0.1, 1]$ window is broad and is not unique to coronal or
jet-base photohadronic scenarios: a range of emission models can
produce $L_\mathrm{hX} / L_\nu$ ratios that fall in this window for the
parameter combinations relevant to flux-limited samples. The
distance-free diagnostic is therefore consistent with, but does not
uniquely select, a photohadronic interpretation.

\subsection{Implications for the Reality of Threshold-Near IceCube
Excesses}

The central observational result of this paper is that the seven
IceCube candidate blazars examined here, all with local pretrial
significances at the $\sim 1$--$2\,\sigma$ level, are
posterior-predictively consistent with the AGN $L_\mathrm{hX}$--$L_\nu$ relation
calibrated independently on the six K2024 sources. This consistency
is non-trivial in the descriptive sense: their hard X-ray luminosities
span more than two orders of magnitude, and the predicted $\log L_\nu$
values vary correspondingly, yet the observed $\log L_\nu$ values
track these predictions with $|z_i| \leq 0.89$ in every case, with
the discriminating power concentrated in the four sources where
$\sigma_{\rm int}$ rather than $\sigma_{\nu,i}$ dominates the
predictive variance.

We emphasize that this result does not, on its own, establish that the
threshold-near IceCube excesses are real astrophysical neutrinos
rather than selected background fluctuations. The null calibration of
Section~\ref{sec:nullcalib} demonstrates that the same $\chi^2_7$
test, when applied to background-only $\hat{n}_s$ realizations
propagated through the same flux pipeline and selection cut, yields
median $\chi^2_{\rm null}$ in the range $2.5$--$2.7$ and median
$p_{\rm null} > 0.9$ across plausible background expectations: the
predictive band is wide enough that both signal and selected
background populate the same low-$\chi^2$, high-$p$ region of the
test distribution. The observed value $\chi^2_7 = 1.58$ sits below
the null medians, in the direction expected if the IceCube excesses
reflect signal, but the gap is not statistically distinguishable from
background within the precision of the present calibration.

What we have therefore established is a conditional consistency
between independent multimessenger information,  measured \nustar luminosities and published IceCube $\hat{n}_s$ values, under the working hypothesis that the AGN hard X-ray--neutrino relation is real and applies to candidate blazars. A detection-level statement against the alternative hypothesis that the threshold-near excesses are background fluctuations requires either (i) an enlarged calibration sample that tightens $\sigma_{\rm int}$ to a value where the test becomes diagnostic of the underlying relation, or (ii) an
X-ray-weighted IceCube stacking likelihood with public event-level data, in which $n_s^{\rm exp}$ is defined a priori from the
$L_\mathrm{hX}$--$L_\nu$ relation and the test statistic is a likelihood ratio
between the signal and background hypotheses rather than a $\chi^2$
against a wide predictive band.

\subsection{Implications for the Neutrino Emission Site}
\label{sec:emissionsite}

Extending the $L_\mathrm{hX}$--$L_\nu$ relation to a pure-blazar test sample
is consistent with, though does not uniquely establish, a unifying
scenario tied to the central engine of AGN. In both blazars and
Seyferts, an SMBH is surrounded by an accretion disk and a hot
corona; in this scenario, the corona supplies the dense, photon-rich
environment that drives $p\gamma$ interactions, and the resulting
electromagnetic cascades from $\gamma$-opaque environments
redistribute energy into the hard X-ray band (this is the cascade
contribution underlying the K2024 relation, not a direct
thermal-Comptonization component). The photon field serving as the $p\gamma$ target need not be the same across classes, the coronal X-ray field in the Seyferts, and more plausibly the BLR UV field in the FSRQs, but in either case the hard X-rays entering the relation are the reprocessed cascade product, which is why a single $L_\mathrm{hX}$--$L_\nu$ relation can span both populations. Direct coronal emission via thermal Comptonization
\citep[e.g.][]{2018ApJS..235....4O} contributes in parallel, and is itself
coupled to the same coronal properties (electron density, temperature,
optical depth) that set the photon-target field for $p\gamma$
interactions, so its luminosity is not statistically independent of
$L_\nu$ either. The cascade contribution, however, provides the more
direct calorimetric link: every $p\gamma$ event reprocesses into hard
X-rays via energy conservation in the $\gamma$-opaque environment,
fixing $L_\mathrm{hX}^{\rm casc} \approx \kappa\,L_\nu$ with $\kappa$ set by
cascade development rather than by independent coronal microphysics.
If the same physical regions drive both, neutrino production becomes
a generic outcome of accretion-powered systems, independent of jet
presence.

Relativistic beaming in blazars enhances detectability and shifts
photon energies into the observed hard X-ray band, but does not by
itself break the $L_\mathrm{hX}$--$L_\nu$ correlation: both luminosities are
beamed by a common Doppler factor $\delta^4$, so the source-frame
ratio $\kappa = L_\mathrm{hX}/L_\nu$ is invariant under beaming. This $\delta$-invariance, and hence the cross-sample comparison of $\kappa$, presumes that the hard X-rays are produced by the same mechanism in both populations, the cascade/coronal channel co-spatial with the neutrino-producing region, so that $L_\mathrm{hX}$ and $L_\nu$ share a common Doppler boost. If part of the blazar 15--55~keV emission arises from jet inverse-Compton scattering that is beamed but not accompanied by neutrino production, $\kappa$ no longer compares the same physical quantity across the two samples. The fact that
both populations cluster within the same $\kappa$ window in the
distance-free diagnostic (Section~\ref{sec:ratio}) is therefore not
trivially expected: if blazar hard X-ray emission were dominated by a
beamed jet component that contributes additional X-rays without a
matching neutrino output, $\kappa_{\rm blazar}$ would be elevated
relative to $\kappa_{\rm Seyfert}$. A proton-loaded jet contributing
to both channels would not produce such a shift, so the observed
agreement is consistent with the corona supplying the dominant photon
target in both populations but does not by itself exclude proton-loaded
jet contributions, given also the broad and selection-biased $\kappa$
window discussed above.

We stress that for FSRQs and BL~Lacs, hard X-rays in the 15--55~keV
band can be produced by jet inverse-Compton emission, by reflected or
coronal emission, or by an electromagnetic cascade contribution
following $\gamma\gamma$ absorption of pionic $\gamma$ rays.
Distinguishing between a coronal and a jet-base origin requires
observations that can separate the two components, simultaneous
broadband SED modeling, Doppler-factor inference, or
correlated-variability studies between hard X-ray and $\gamma$-ray
emission. For \txs, the anti-correlation between \fermi-LAT and
\nustar light curves reported in K2024 provides circumstantial
evidence against jet dominance of the 15--55~keV band in that source,
but a generalization to the rest of the blazar sample requires
source-by-source SED decomposition that is beyond the present scope. The observed consistency therefore constrains the
physical conditions in the emission region (photon energy, density,
compactness) but does not by itself uniquely determine the emission
site.

\subsection{Caveats and Limitations}
\label{sec:caveats}
\textit{Distance and selection biases in the secondary permutation
test.} The flux-space permutation test (Section~\ref{sec:permutation})
preserves the $d_L^{\,2}$ coupling identically in observation and
null, so the distance bias is corrected by construction; the
luminosity-space variant (5.07$\,\sigma$) is shown only to quantify
this contribution. The IceCube selection bias acts conservatively in
the flux-space test: because shuffled neutrino fluxes are drawn from
a neutrino-bright pool, the null already exhibits less scatter than
a truly random blazar sample would, so the reported $3.23\,\sigma$ is
a lower bound on the significance achievable with a blind blazar
control. Such a control cannot be constructed from publicly available
IceCube data alone; it would require source-specific neutrino flux
estimates for an $N \sim 10^2$ blazar sample not pre-selected as
neutrino candidates.

A complementary analysis by \citet{LuoLuLiang2026} 
using a 1114-AGN Swift-BAT sample and 10 years of public IceCube data
demonstrates explicitly that the $L_X$--$L_\nu$ correlation can be
reproduced from random sky positions through the same $d_L^2$
coupling that motivates our flux-space permutation test
(Section~\ref{sec:permutation}). Their Monte Carlo simulations
reproduce $r \approx 0.85$--$0.93$ in luminosity space at TS cuts
matching the K2024 sample, but yield $r_F = 0.07$--$0.20$ in
flux space. Our framing as a
conditional consistency check, together with the flux-space
permutation test and the distance-free $\kappa$ diagnostic of
Section~\ref{sec:ratio}, is designed to be robust exactly against the
selection-effect concern the authors raised.

\textit{Neutrino luminosity in the 0.3-100 TeV energy range.} $L_\nu$ is luminosity integrated over 0.3--100~TeV (Section~\ref{sec:nuflux}); for FSRQs whose physical neutrino SED may peak at $\gtrsim$PeV via BLR target $p\gamma$ \citep{2014JHEAp...3...29D}, it is bound-specific luminosity, a lower bound on the bolometric neutrino luminosity. Because the band is applied identically to the calibration and test samples and the relation is calorimetric (the hard X-rays trace the cascade product; Section~\ref{sec:emissionsite}), the band choice does not bias the comparison. Extending the integration to 10~PeV (Appendix~\ref{app:robustness}) shifts only the hardest-spectrum (lowest-$\hat{n}_s$) sources and leaves the consistency conclusion unchanged: under the plug-in $\sigma_{\rm pred}$, $\chi^2_7 = 2.46$ ($p = 0.93$); once the relation-parameter covariance is included, the band choice becomes immaterial ($\chi^2_7 = 0.78$ and $0.96$ for the 0.3--100~TeV and 0.3--10~PeV bands, respectively).

\textit{Non-simultaneity of \nustar and IceCube data}. The
IceCube data underlying the published $\hat{n}_s$ and $\hat{\gamma}_\nu$
values \citep{ngc1068_2022_dataset} span 13~May 2011 to 29~May 2020,
while several of our \nustar epochs (Table~\ref{tab:nustar_results})
post-date this window. For a strict source-frame $L_\mathrm{hX}$--$L_\nu$
correlation in which $L_\mathrm{hX}$ traces the photon field driving $p\gamma$
neutrino production at the moment of emission, comparing a post-2020
\nustar flux to an IceCube excess accumulated over 2011--2020 is
physically imperfect. However, the \nustar flux variability across
multi-epoch sources is typically within a factor of $\sim$2--3
(Section~\ref{sec:xray}; $\sim 0.3$--$0.5$\,dex), smaller than the
intrinsic scatter $\sigma_{\rm int} = 0.60$\,dex that sets the
predictive band, so the consistency test is not driven by
epoch-to-epoch flux variability. We have verified that excluding the
post-2020 \nustar epochs and using only contemporaneous observations
does not change the qualitative outcome (the four high-$\hat{n}_s$
sources retain $|z| < 1$), but the impact is non-negligible at the
source level for the most variable members of the sample, and
time-resolved follow-up remains a natural improvement for individual
blazar studies.

\textit{Asymmetric error treatment between calibration and test
samples and regime mismatch of the empirical scaling factor.} The
calibration uses spectrum-derived $\sigma_{\nu,i}$ from the IceCube
profile likelihood (on average $\sim 4.5\times$ the Poisson form);
the seven test sources lack a public profile and use the scaled
Poisson form (Section~\ref{sec:nuflux}). The K2024 calibration spans
$\hat{n}_s \in [5, 79]$; three test sources have $\hat{n}_s \leq 3$
where the Poisson form is less accurate. The high-$\hat{n}_s$
vs.\ low-$\hat{n}_s$ split (Section~\ref{sec:postpred}) addresses
this: the four high-$\hat{n}_s$ sources give $\chi^2_4 = 1.24$; the
three low-$\hat{n}_s$ sources are $\sigma_{\nu,i}$-dominated. A
sensitivity check reducing the scaling factor to $2.5$ (below the
calibrated range $3.7$--$5.2$) yields $\chi^2_7 = 2.63$ and
$\max|z_i| = 1.02$, all within $|z| < 1.1$, robust to the
scaling-factor choice.

\textit{Limited statistical power of the consistency test.}
The intrinsic scatter recovered from the calibration is
$\sigma_{\rm int} = 0.60^{+0.78}_{-0.33}$\,dex, broad relative to the
typical predictive residual. Adopted at its posterior median in the
plug-in formulation of Section~\ref{sec:postpred}, this is the
dominant component of $\sigma_{{\rm pred},i}$ for the high-$\hat{n}_s$
test sources and reflects the limited calibration sample size
($N_{\rm cal} = 6$). The $\chi^2_7 = 1.58$ ($p = 0.980$) should
therefore be read as ``no source is in tension with the predictive
band, even where the band is tightest,'' rather than as a positive
measurement of relation tightness beyond the prior. The null
calibration of Section~\ref{sec:nullcalib} quantifies this: median
$\chi^2_{\rm null}$ under background-only realizations is $\sim 2.5$,
showing that a low $\chi^2_7$ value is the generic outcome of the
present pipeline rather than a signature of signal. Full
marginalization over the calibration chain inflates
$\sigma_{{\rm pred},i}$ to $\gtrsim 2$\,dex per source via the heavy
upper tail of the $\sigma_{\rm int}$ posterior at $N_{\rm cal} = 6$,
which motivates the plug-in choice (Section~\ref{sec:postpred}).

\section{Conclusion}
\label{sec:conclusion}

We have addressed the question of whether threshold-near IceCube
excesses associated with seven \nustar-observed candidate blazars
are statistically consistent with being drawn from the hard
X-ray--neutrino relation reported by K2024 for the six AGN with the
highest IceCube significances, under the working hypothesis that the
published $\hat{n}_s$ values reflect signal. We tested the new sample
against the K2024 correlation using a posterior-predictive
consistency check, characterized the test's discriminating power as a
function of $\hat{n}_s$, and calibrated it against an explicit
background-only null.

The central findings are:
\begin{enumerate}
\item The $L_\mathrm{hX}$--$L_\nu$ relation calibrated on the six K2024 sources
only, using a Bayesian regression with errors on both axes
\citep[\texttt{linmix} implementation of][]{Kelly2007}, gives $\beta =
1.26^{+0.40}_{-0.38}$ (consistent with $\beta = 1$ at $\sim
0.7\,\sigma$) and $\sigma_{\rm int} = 0.60^{+0.78}_{-0.33}$\,dex.
The credible intervals are wide given $N_{\rm cal} = 6$.
\item All seven new blazars are posterior-predictively consistent
with the calibration: $|z_i| \leq 0.89$ for every source, global
$\chi^2_7 = 1.58$, $p = 0.980$ (see Fig. \ref{fig:postpred}). The discriminating power of the test
is concentrated in the four high-$\hat{n}_s$ sources ($\hat{n}_s \geq
9$), which give $\chi^2_4 = 1.24$ ($p = 0.87$); the three
low-$\hat{n}_s$ sources ($\hat{n}_s \leq 3$) give $\chi^2_3 = 0.34$
($p = 0.95$) and lie in the regime where $\sigma_{\nu,i}$ dominates
the predictive variance, rendering the test insensitive there.
\item A null calibration with background-only $\hat{n}_s$
realizations propagated through the same pipeline gives median
$\chi^2_{\rm null}$ in the range $2.48$--$2.74$ across plausible
background expectations $\lambda_{\rm bg} \in [0.3, 2.0]$, with
$P(p_{\rm null} > 0.5) = 1.00$ in every case (see Fig. \ref{fig:null}). The $\chi^2_7$ test as
constructed therefore does not, on its own, distinguish between the
signal and selected-background hypotheses; the observed value sits
below the null median by $\sim 1$ unit, in the direction expected for
signal but not statistically distinguishable from background at the
present calibration precision.
\item The distance-free $\log(L_\mathrm{hX}/L_\nu)$ diagnostic places both
populations within the photohadronic $\kappa \in [0.1,1]$ band (see Fig. \ref{fig:ratio}), with
calibration and test means consistent with each other at $\sim
0.7\,\sigma$. We caution that the joint \nustar$+$IceCube selection
envelope preferentially populates this same $\kappa$ window, so the
agreement is necessary but not sufficient evidence of common physics.
\item As a finite-sample diagnostic, a weighted flux-space
permutation test on the joint 13-source sample gives
$\sigma_{\rm scat}^{\rm w,\,obs} = 0.321$\,dex versus a null mean of
$1.08$\,dex; $p = 6.3\times10^{-4}$ ($3.23\,\sigma$), $d_L^2$-bias
controlled by construction (see Fig. \ref{fig:permutation}). This test probes the relative ordering
of $L_\mathrm{hX}$ and $L_\nu$ within the joint sample and is partially
independent of the predictive-band-dominated $\chi^2$ test.
\end{enumerate}

We interpret these results as a conditional consistency check:
under the working assumption that the IceCube excesses reflect signal,
the seven candidate blazars are statistically indistinguishable from
the K2024 calibration in posterior-predictive residual, in
distance-free luminosity ratio, and in flux-space permutation
structure. The analysis is not an independent IceCube detection claim,
and the present test does not, on its own, adjudicate between the
signal hypothesis and the alternative that the threshold-near
$\hat{n}_s$ values are selected background fluctuations: the
predictive band is wide enough that both populate the same region of
the $\chi^2$ distribution. A detection-level statement for the
candidate-blazar population would require either (i) an enlarged
calibration sample that tightens $\sigma_{\rm int}$ to a value where
the consistency test becomes diagnostic, or (ii) an X-ray-weighted
IceCube stacking likelihood with public event-level data, defined a
priori from the $L_\mathrm{hX}$--$L_\nu$ relation. We identify the latter as
the natural next step.

Future work will benefit from (i) an enlarged calibration sample to
sharpen $\sigma_{\rm int}$ and the posterior-predictive band,
(ii) targeted \nustar observations of additional IceCube candidate
blazars currently lacking hard X-ray coverage, ideally contemporaneous
with future IceCube exposure windows to address the non-simultaneity
caveat,
(iii) public release of IceCube point-source likelihood profiles (or
a per-source analogue) for the threshold-near sample, which would
replace the $\sim 4.5\times$ empirical scaling of the Poisson
uncertainty with first-principles $\sigma_{\nu,i}$ values, especially
in the low-$\hat{n}_s$ regime where the present scaling is least
calibrated,
(iv) a blind blazar control sample to quantify selection effects
directly, and (v) collaboration with IceCube on a model-weighted
stacking analysis with $n_s^{\rm exp}$ defined a priori from the
$L_\mathrm{hX}$--$L_\nu$ relation.

\begin{acknowledgments}
We thank Ceaser C.\ Stringfield for carrying out the \nustar spectral
analyzes of blazars in our sample, and Kaya Mori and Jooyun Woo for
their technical and managerial support of that work. E.K.\ thanks
funding from the NKFIH excellence grant TKP2021-NKTA-64. J.B.T.,
E.K., and A.F.\ acknowledge support from the German Science Foundation
DFG, via the Collaborative Research Center \textit{SFB1491: Cosmic
Interacting Matters -- from Source to Signal} (grant no.\ 445052434).
Francis Halzen is funded by the U.S. National Science Foundation under
grants PHY-2209445 and OPP-2042807. C.R. acknowledges support from
SNSF Consolidator grant F01$-$13252. 
SdP acknowledges support from ERC Advanced Grant 789410. 
This research has made use of the NuSTAR Data Analysis Software
(NuSTARDAS) jointly developed by the ASI Space Science Data Center
(SSDC, Italy) and the California Institute of Technology (Caltech,
USA). This work made use of data supplied by the UK Swift Science Data
Centre at the University of Leicester. This paper makes use of
publicly available Fermi-LAT data provided online by the
\url{https://fermi.gsfc.nasa.gov/ssc/data/access/} Fermi Science
Support Center. This work made use of
Astropy\footnote{\url{http://www.astropy.org}}: a community-developed
core Python package and an ecosystem of tools and resources for
astronomy \citep{2013A&A...558A..33A,2018AJ....156..123A,
2022ApJ...935..167A}, and the \texttt{linmix}
package by J.~Meyers \citep{Kelly2007}.
\end{acknowledgments}

\facilities{IceCube, \nustar, \fermi (LAT)}

\software{astropy \citep{2013A&A...558A..33A,2018AJ....156..123A,
2022ApJ...935..167A},
\texttt{linmix} (J.~Meyers, port of \citealt{Kelly2007}),
NuSTARDAS, XSPEC}

\clearpage
\appendix

\section{Multiwavelength Light Curves}
\label{app:multiwave}
\begin{figure*}[h!]
\centering
\includegraphics[width=0.85\textwidth]{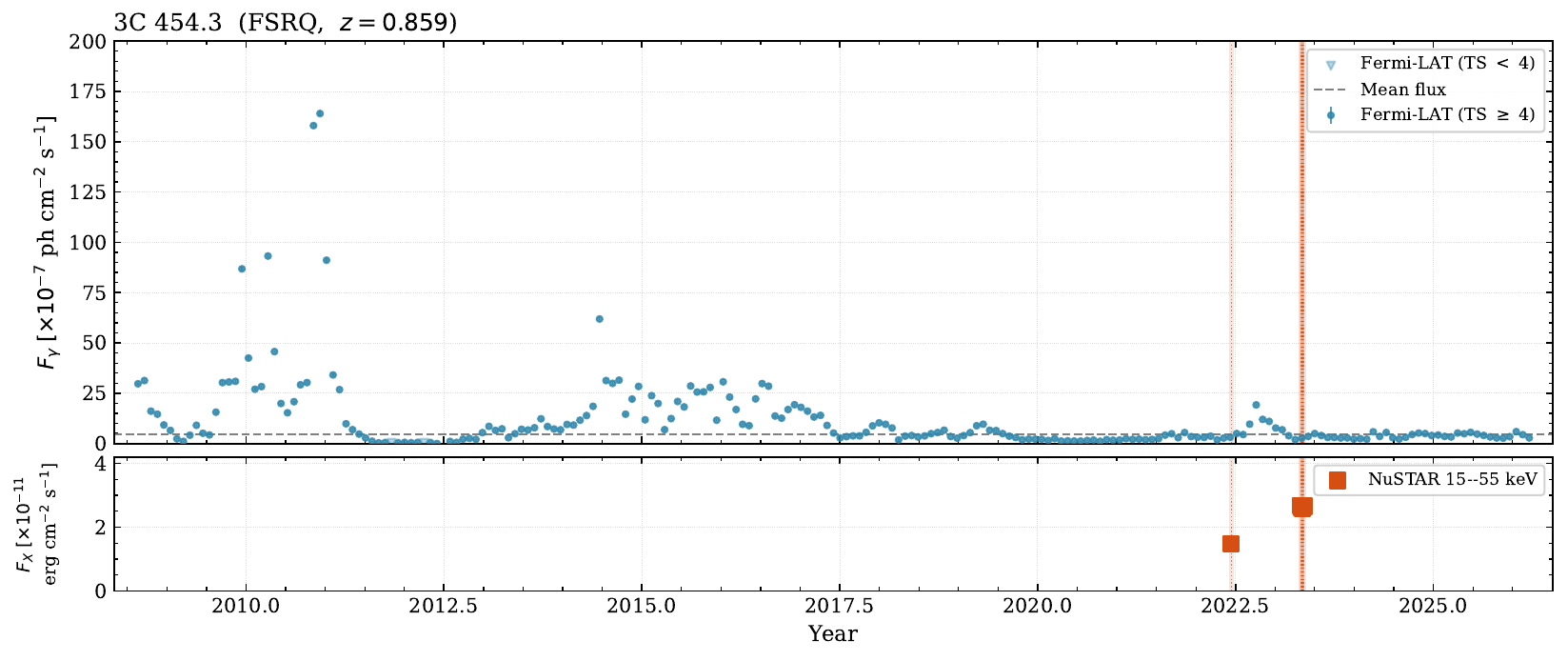}\\[1ex]
\includegraphics[width=0.85\textwidth]{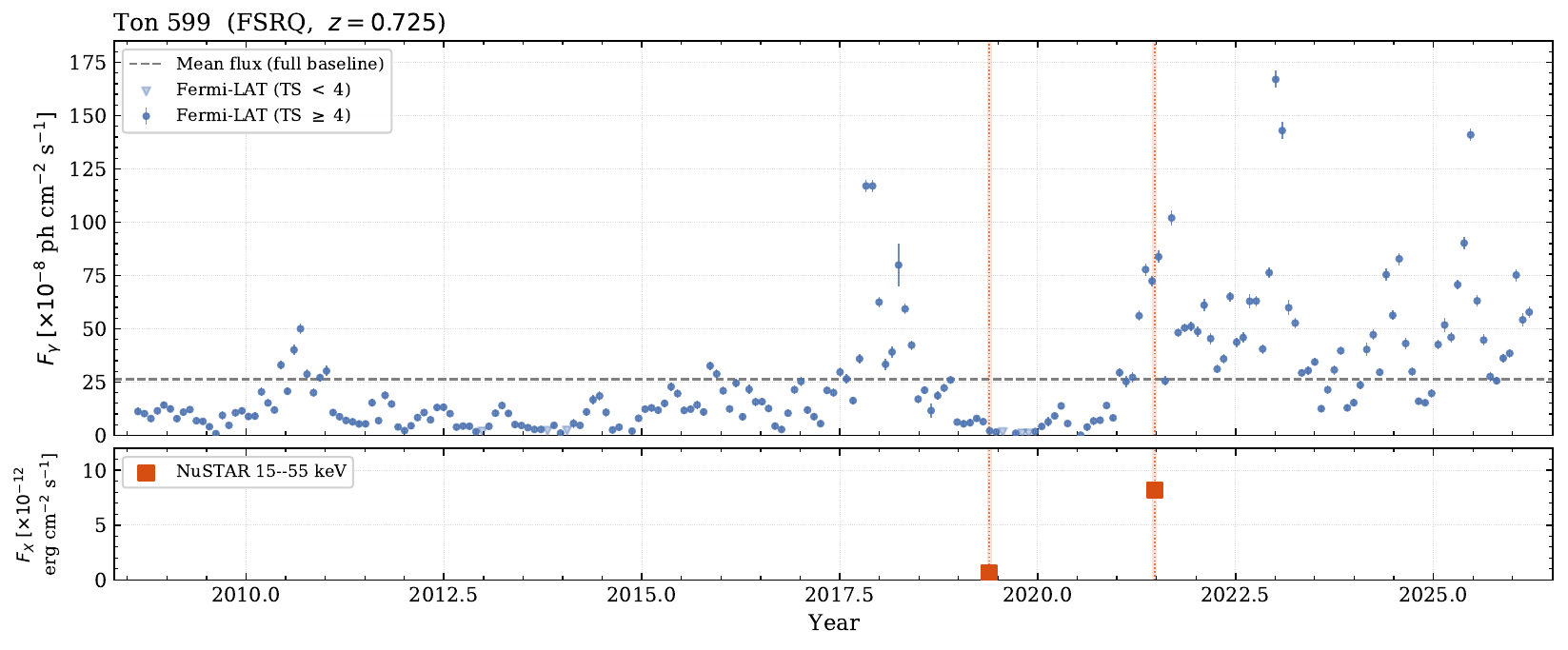}
\caption{Top: Monthly \fermi-LAT 0.1--100~GeV photon flux (top) and
\nustar 15--55~keV energy flux (bottom) of 3C~454.3 as a function
of time. Filled circles show \fermi-LAT detections (TS~$\geq 4$);
downward triangles indicate upper limits (TS~$< 4$). The dashed line
marks the quiescent mean flux over 2017--2022. Orange shaded bands and
dotted vertical lines indicate the epochs of \nustar observations. Bottom: Monthly \fermi-LAT 0.1--100~GeV photon flux (top) and
\nustar 15--55~keV energy flux (bottom) Ton~599 (FSRQ,
$z = 0.725$). Symbols and lines as in the top panel. Light curves from \citet{lcr}. }
\label{fig:3c454_lc}
\end{figure*}

\begin{figure*}
\centering
\includegraphics[width=0.85\textwidth]{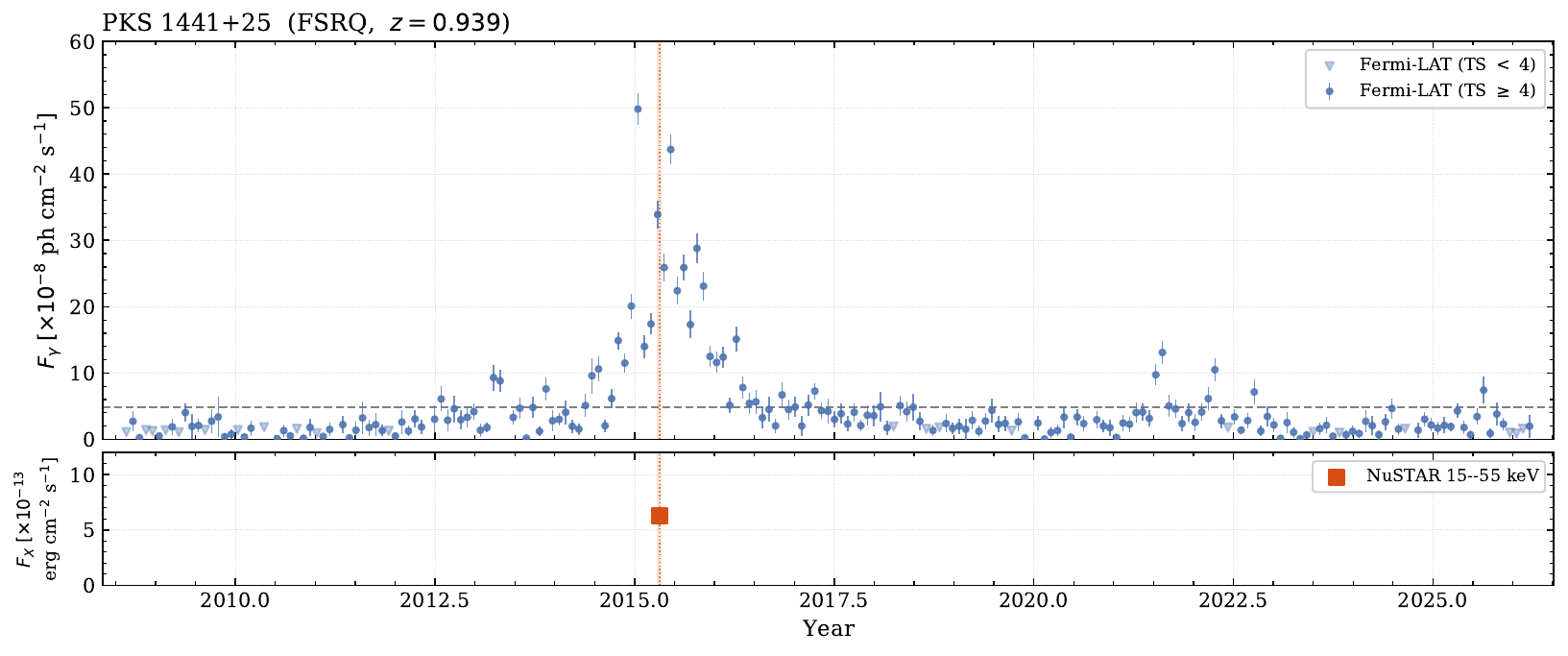}\\[1ex]
\includegraphics[width=0.85\textwidth]{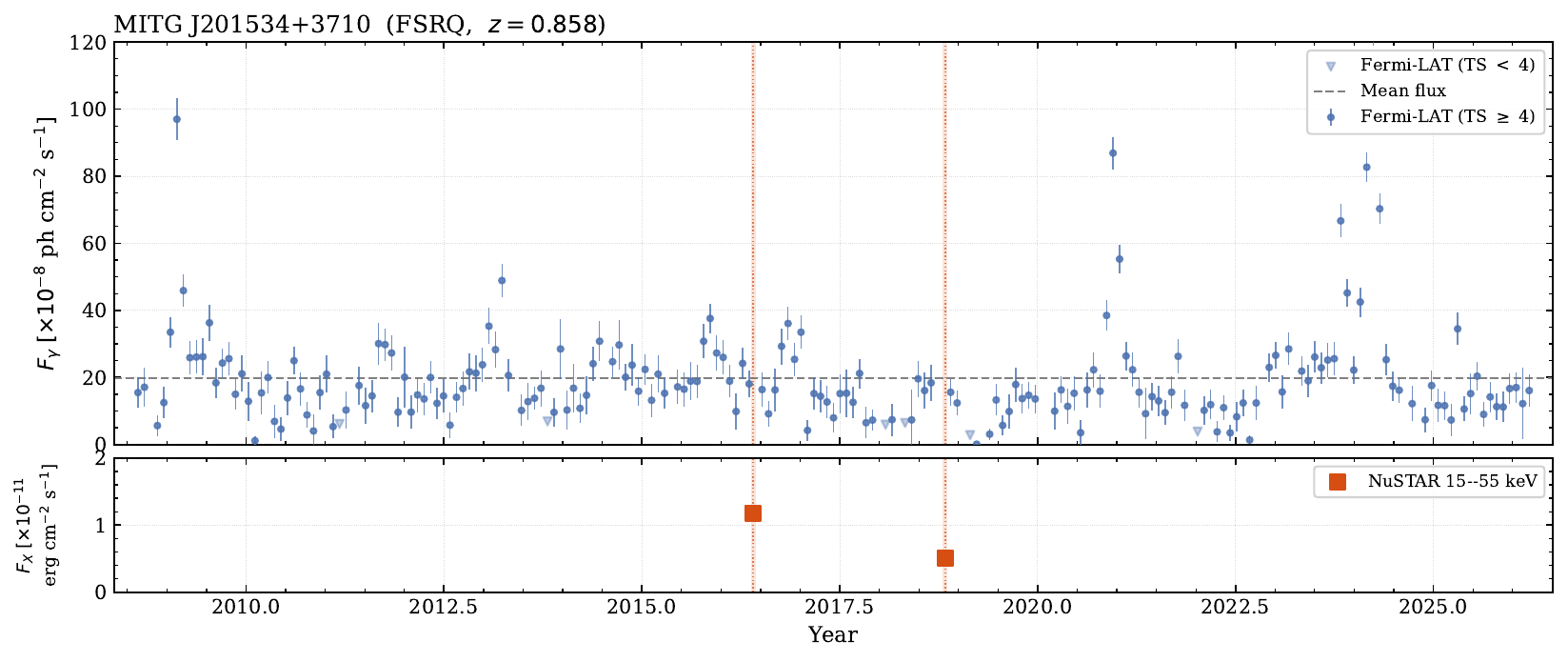}\\[1ex]
\includegraphics[width=0.85\textwidth]{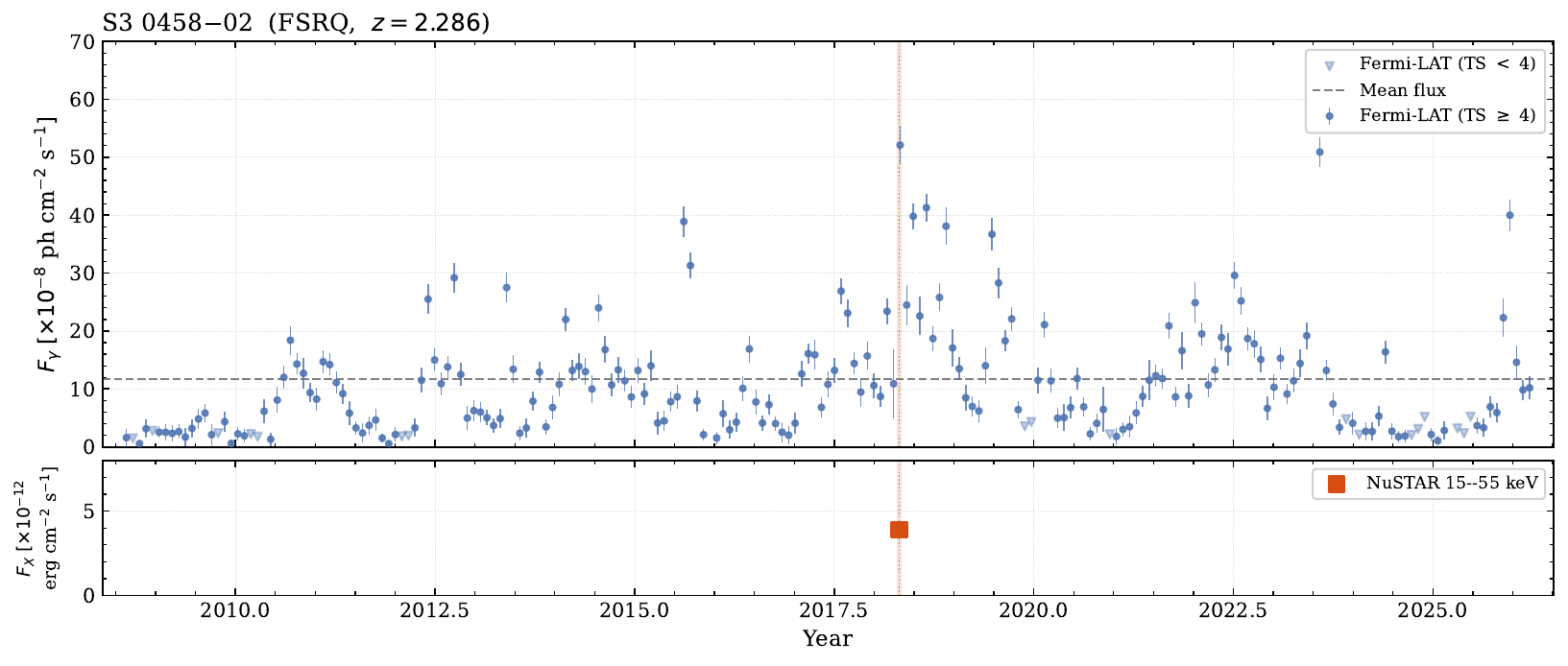}
\caption{\fermi-LAT and \nustar light curves of, from top to bottom:
PKS~1441+25 (FSRQ, $z = 0.939$; the neutrino spectral index was fitted
freely); MITG~J201534+3710 (FSRQ, $z = 0.858$); and S3~0458$-$02
(FSRQ, $z = 2.286$). Symbols and lines as in
Figure~\ref{fig:3c454_lc}. Light curves from \citet{lcr}.}
\label{fig:lc_group2}
\end{figure*}

\begin{figure*}
\centering
\includegraphics[width=0.85\textwidth]{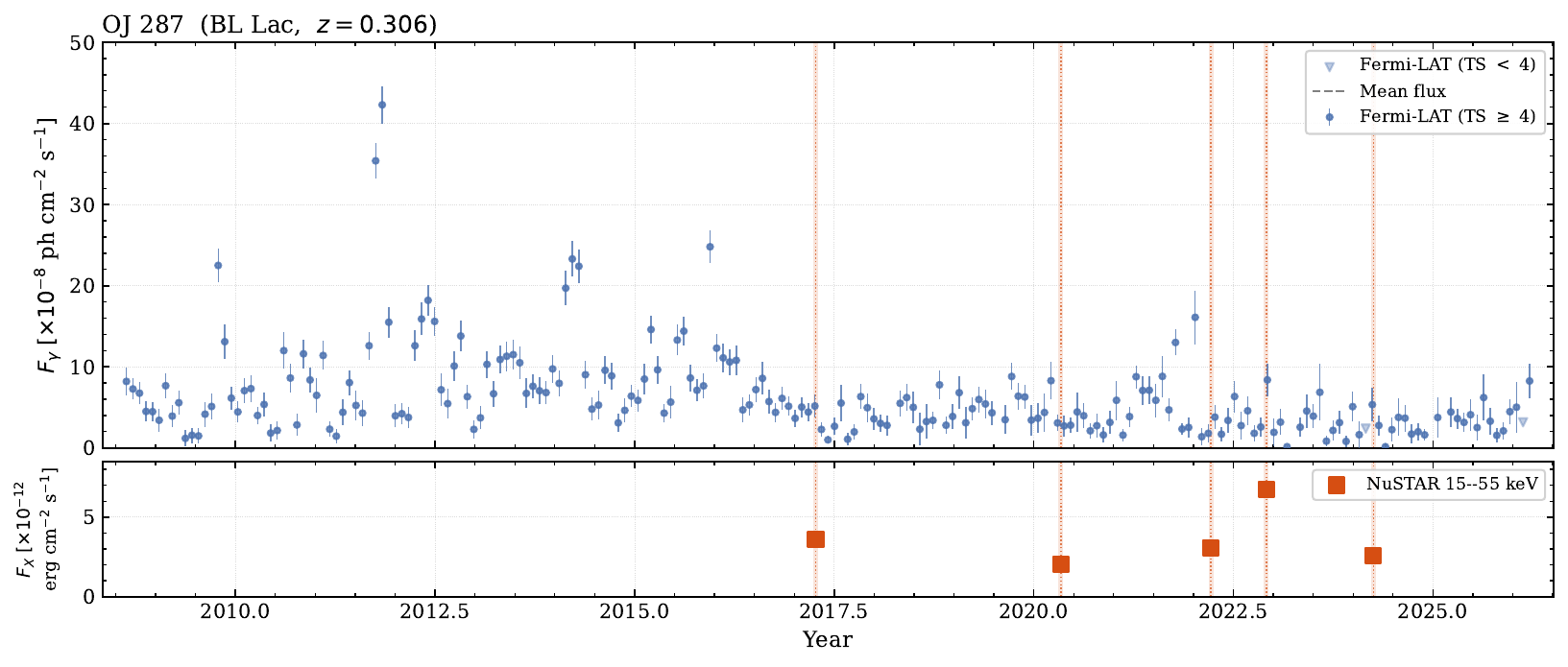}\\[1ex]
\includegraphics[width=0.85\textwidth]{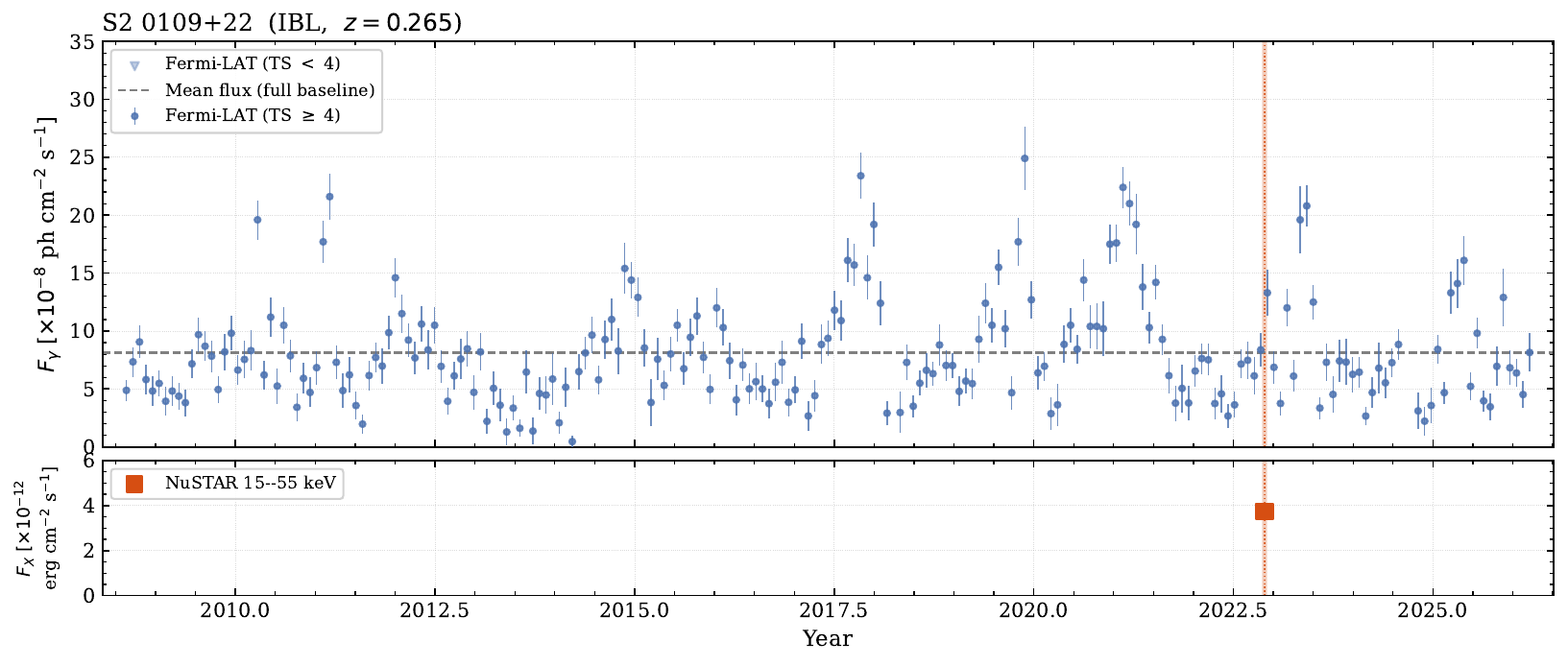}
\caption{\fermi-LAT and \nustar light curves of, from top to bottom:
OJ~287 (BL~Lac, $z = 0.306$); S2~0109+22 (IBL, $z = 0.265$; the
neutrino spectral index was fitted freely). Symbols and lines as in Figure~\ref{fig:3c454_lc}.
Light curves from \citet{lcr}.}
\label{fig:lc_group3}
\end{figure*}

Figures~\ref{fig:3c454_lc}--\ref{fig:lc_group3} show the monthly
\fermi-LAT 0.1--100~GeV photon flux (top panel) and \nustar 15--55~keV
energy flux (bottom panel) for each source in the blazar sample.
Filled circles indicate \fermi-LAT detections with test statistic
TS~$\geq 4$; downward triangles indicate monthly upper limits
(TS~$< 4$). The dashed horizontal line marks the quiescent mean flux
computed over the full baseline (2008--2025) or over the 2017--2022
period for highly variable sources. Orange shaded bands and dotted
vertical lines indicate the epochs of individual \nustar observations.

\clearpage
\section{Robustness: Neutrino Energy Band and Predictive-Variance Prescriptions}
\label{app:robustness}

This appendix quantifies the two robustness checks referenced in the main text: extending the neutrino integration band from 0.3--100~TeV to 0.3--10~PeV, and propagating the relation-parameter $(\alpha,\beta)$ covariance into the predictive variance.

\textit{Energy-band extension}. Since the band-integrated energy flux of a power law $E\phi\propto E^{1-\gamma_\nu}$ enters $L_\nu$ multiplicatively, raising the upper bound from 100~TeV to 10~PeV rescales $L_\nu$ by a factor that depends only on $\gamma_\nu$ (the normalization $\phi_0$, the redshift, and the $K$-correction all cancel). For the soft spectra of our sample the shift is negligible except for the hardest-spectrum sources: in the calibration only TXS~0506+056 ($\gamma_\nu=2$) shifts appreciably ($+0.25$~dex), and in the test sample the shifts are concentrated in the two lowest-$\hat{n}_s$ sources (Table~\ref{tab:robustness}). Re-running the full calibration and consistency test over the extended band steepens the relation only marginally ($\beta: 1.26\to1.30$, within the $\pm0.40$ credible width) and leaves all sources consistent.

\textit{Predictive-variance prescriptions.} Naively marginalizing over the full $(\alpha,\beta)$ posterior is uninformative ($\sigma_{{\rm pred},i}=2$--$5$\,dex, $\chi^2_7=0.15$): the chain standard deviation of $\beta$ is $0.65$--$1.26$ (depending on the band), a factor $\sim 2$--$3$ larger than the 68\% credible half-width ($\approx0.38$), so rare steep-slope samples extrapolated $1$--$3$\,dex beyond the calibration range dominate $\sigma_{\rm pred}$, the same heavy-tail behaviour at $N_{\rm cal}=6$ that motivates the plug-in treatment of $\sigma_{\rm int}$ (Section~\ref{sec:postpred}). We therefore propagate the parameter uncertainty using robust (16--84 percentile) scales in a pivot parametrization at $\log L_\mathrm{hX}=43.39$, where intercept and slope are essentially uncorrelated (corr $\approx -0.04$ to $-0.08$): $\mathrm{sd}(\alpha_{\rm pivot})\approx0.33$, $\mathrm{sd}(\beta)\approx0.38$, with
\begin{equation}
\sigma_{{\rm pred},i}^2 = \sigma_{\nu,i}^2 + \sigma_{\rm int,med}^2 + \beta_{\rm med}^2\sigma_{X,i}^2 + \left[\sigma_{\alpha,\rm piv}^2 + (\log L_{\mathrm{hX},i} - 43.39)^2\,\sigma_\beta^2\right].
\label{eq:sigmapred_cov}
\end{equation}
With this term, $\sigma_{{\rm pred},i}$ at the high-luminosity sources grows from $\sim0.7$ to $1.4$--$2.4$\,dex (Table~\ref{tab:robustness}), confirming that the plug-in predictions are tight there; the predictive $z$-scores roughly halve. The sample remains consistent in every prescription and band (Table~\ref{tab:robustness_chi2}), and the robust-covariance treatment removes essentially all sensitivity of the result to the band choice.

\begin{deluxetable*}{lccccccc}
\tablecaption{Robustness of the consistency test to the neutrino energy band and to the predictive-variance prescription, for the seven test sources. $\log L_\nu$ is given for the primary 0.3--100~TeV band and the extended 0.3--10~PeV band ($\Delta$ their difference); $\sigma_{\rm pred}$ and $z$ are for the plug-in (Eq.~\ref{eq:sigmapred}) and robust-covariance (Eq.~\ref{eq:sigmapred_cov}) prescriptions at the primary band.\label{tab:robustness}}
\tablehead{
\colhead{Source} & \colhead{$\log L_\nu^{100\,{\rm TeV}}$} & \colhead{$\log L_\nu^{10\,{\rm PeV}}$} & \colhead{$\Delta$} & \colhead{$\sigma_{\rm pred}^{\rm plug}$} & \colhead{$z^{\rm plug}$} & \colhead{$\sigma_{\rm pred}^{\rm rob}$} & \colhead{$z^{\rm rob}$}
}
\startdata
3C 454.3            & 46.65 & 47.68 & $+1.02$ & 2.04 & $-0.55$ & 2.46 & $-0.46$ \\
S2 0109+22          & 44.82 & 44.83 & $+0.01$ & 0.87 & $-0.60$ & 1.09 & $-0.50$ \\
PKS 1441+25         & 46.29 & 46.46 & $+0.17$ & 1.29 & $+0.13$ & 1.56 & $+0.10$ \\
MITG J201534+3710   & 46.57 & 46.57 & $0.00$  & 0.75 & $-0.89$ & 1.42 & $-0.49$ \\
S3 0458$-$02        & 48.23 & 48.23 & $0.00$  & 0.89 & $+0.14$ & 1.71 & $+0.06$ \\
OJ 287              & 45.37 & 45.37 & $0.00$  & 0.78 & $-0.26$ & 1.06 & $-0.21$ \\
Ton 599             & 46.45 & 46.45 & $0.00$  & 1.51 & $-0.15$ & 1.83 & $-0.13$ \\
\enddata
\end{deluxetable*}

\begin{table}
\centering
\footnotesize
\caption{\textbf{Global $\chi^2_7$ ($p$-value) under three predictive-variance prescriptions and two integration bands.}\label{tab:robustness_chi2}}
\begin{tabular}{lcc}
\hline\hline
Prescription & 0.3--100~TeV & 0.3--10~PeV \\
\hline
Plug-in (Eq.~\ref{eq:sigmapred})       & 1.58 (0.980) & 2.46 (0.930) \\
Robust $(\alpha,\beta)$ cov.\ (Eq.~\ref{eq:sigmapred_cov}) & 0.78 (0.998) & 0.96 (0.995) \\
Full marg.\ & \multicolumn{2}{c}{uninformative ($\sigma_{\rm pred}\!=\!2$--$5$\,dex; $\chi^2_7\!=\!0.15$ at 10~PeV)} \\
\hline
\end{tabular}
\end{table}


\begin{thebibliography}{}
\expandafter\ifx\csname natexlab\endcsname\relax\def\natexlab#1{#1}\fi
\providecommand{\url}[1]{\href{#1}{#1}}
\providecommand{\dodoi}[1]{doi:~\href{http://doi.org/#1}{\nolinkurl{#1}}}
\providecommand{\doeprint}[1]{\href{http://ascl.net/#1}{\nolinkurl{http://ascl.net/#1}}}
\providecommand{\doarXiv}[1]{\href{https://arxiv.org/abs/#1}{\nolinkurl{https://arxiv.org/abs/#1}}}

\bibitem[{M.~G. {Aartsen} {et~al.}(2013){Aartsen}, {Abbasi}, {Abdou}, {Ackermann}, {Adams}, {et~al.}}]{2013PhRvL.111b1103A}
{Aartsen}, M.~G., {Abbasi}, R., {Abdou}, Y., {et~al.} 2013, \bibinfo{title}{{First Observation of PeV-Energy Neutrinos with IceCube},} \prl, 111, 021103, \dodoi{10.1103/PhysRevLett.111.021103}

\bibitem[{M.~G. {Aartsen} {et~al.}(2020){Aartsen}, {Ackermann}, {Adams}, {Aguilar}, {Ahlers}, {Ahrens}, {Alispach}, {Andeen}, {Anderson}, {Ansseau}, {Anton}, {Arg{\"u}elles}, {Auffenberg}, {Axani}, {Backes}, {Bagherpour}, {Bai}, {Balagopal}, {Barbano}, {Barwick}, {Bastian}, {Baum}, {Baur}, {Bay}, {Beatty}, {Becker}, {Becker Tjus}, {BenZvi}, {Berley}, {Bernardini}, {Besson}, {Binder}, {Bindig}, {Blaufuss}, {Blot}, {Bohm}, {B{\"o}rner}, {B{\"o}ser}, {Botner}, {B{\"o}ttcher}, {Bourbeau}, {Bourbeau}, {Bradascio}, {Braun}, {Bron}, {Brostean-Kaiser}, {Burgman}, {Buscher}, {Busse}, {Carver}, {Chen}, {Cheung}, {Chirkin}, {Choi}, {Clark}, {Classen}, {Coleman}, {Collin}, {Conrad}, {Coppin}, {Correa}, {Cowen}, {Cross}, {Dave}, {De Clercq}, {DeLaunay}, {Dembinski}, {Deoskar}, {De Ridder}, {Desiati}, {de Vries}, {de Wasseige}, {de With}, {DeYoung}, {Diaz}, {D{\'\i}az-V{\'e}lez}, {Dujmovic}, {Dunkman}, {Dvorak}, {Eberhardt}, {Ehrhardt}, {Eller}, {Engel}, {Evenson}, {Fahey}, {Fazely}, {Felde}, {Filimonov}, {Finley}, {Fox},
  {Franckowiak}, {Friedman}, {Fritz}, {Gaisser}, {Gallagher}, {Ganster}, {Garrappa}, {Gerhardt}, {Ghorbani}, {Glauch}, {Gl{\"u}senkamp}, {Goldschmidt}, {Gonzalez}, {Grant}, {Griffith}, {Griswold}, {G{\"u}nder}, {G{\"u}nd{\"u}z}, {Haack}, {Hallgren}, {Halliday}, {Halve}, {Halzen}, {Hanson}, {Haungs}, {Hebecker}, {Heereman}, {Heix}, {Helbing}, {Hellauer}, {Henningsen}, {Hickford}, {Hignight}, {Hill}, {Hoffman}, {Hoffmann}, {Hoinka}, {Hokanson-Fasig}, {Hoshina}, {Huang}, {Huber}, {Huber}, {Hultqvist}, {H{\"u}nnefeld}, {Hussain}, {In}, {Iovine}, {Ishihara}, {Japaridze}, {Jeong}, {Jero}, {Jones}, {Jonske}, {Joppe}, {Kang}, {Kang}, {Kappes}, {Kappesser}, {Karg}, {Karl}, {Karle}, {Katz}, {Kauer}, {Kelley}, {Kheirandish}, {Kim}, {Kintscher}, {Kiryluk}, {Kittler}, {Klein}, {Koirala}, {Kolanoski}, {K{\"o}pke}, {Kopper}, {Kopper}, {Koskinen}, {Kowalski}, {Krings}, {Kr{\"u}ckl}, {Kulacz}, {Kurahashi}, {Kyriacou}, {Labare}, {Lanfranchi}, {Larson}, {Lauber}, {Lazar}, {Leonard}, {Leszczy{\'n}ska}, {Leuermann}, {Liu},
  {Lohfink}, {Lozano Mariscal}, {Lu}, {Lucarelli}, {L{\"u}nemann}, {Luszczak}, {Lyu}, {Ma}, {Madsen}, {Maggi}, {Mahn}, {Makino}, {Mallik}, {Mallot}, {Mancina}, {Mari{\c{s}}}, {Maruyama}, {Mase}, {Matis}, {Maunu}, {McNally}, {Meagher}, {Medici}, {Medina}, {Meier}, {Meighen-Berger}, {Menne}, {Merino}, {Meures}, {Micallef}, {Mockler}, {Moment{\'e}}, {Montaruli}, {Moore}, {Morse}, {Moulai}, {Muth}, {Nagai}, {Naumann}, {Neer}, {Niederhausen}, {Nisa}, {Nowicki}, {Nygren}, {Obertacke Pollmann}, {Oehler}, {Olivas}, {O'Murchadha}, {O'Sullivan}, {Palczewski}, {Pandya}, {Pankova}, {Park}, {Peiffer}, {P{\'e}rez de los Heros}, {Philippen}, {Pieloth}, {Pinat}, {Pizzuto}, {Plum}, {Porcelli}, {Price}, {Przybylski}, {Raab}, {Raissi}, {Rameez}, {Rauch}, {Rawlins}, {Rea}, {Reimann}, {Relethford}, {Renschler}, {Renzi}, {Resconi}, {Rhode}, {Richman}, {Robertson}, {Rongen}, {Rott}, {Ruhe}, {Ryckbosch}, {Rysewyk}, {Safa}, {Sanchez Herrera}, {Sandrock}, {Sandroos}, {Santander}, {Sarkar}, {Sarkar}, {Satalecka}, {Schaufel},
  {Schieler}, {Schlunder}, {Schmidt}, {Schneider}, {Schneider}, {Schr{\"o}der}, {Schumacher}, {Sclafani}, {Seckel}, {Seunarine}, {Shefali}, {Silva}, {Snihur}, {Soedingrekso}, {Soldin}, {Song}, {Spiczak}, {Spiering}, {Stachurska}, {Stamatikos}, {Stanev}, {Stein}, {Steinm{\"u}ller}, {Stettner}, {Steuer}, {Stezelberger}, {Stokstad}, {St{\"o}{\ss}l}, {Strotjohann}, {St{\"u}rwald}, {Stuttard}, {Sullivan}, {Taboada}, {Tenholt}, {Ter-Antonyan}, {Terliuk}, {Tilav}, {Tollefson}, {Tomankova}, {T{\"o}nnis}, {Toscano}, {Tosi}, {Trettin}, {Tselengidou}, {Tung}, {Turcati}, {Turcotte}, {Turley}, {Ty}, {Unger}, {Unland Elorrieta}, {Usner}, {Vandenbroucke}, {Van Driessche}, {van Eijk}, {van Eijndhoven}, {Vanheule}, {van Santen}, {Vraeghe}, {Walck}, {Wallace}, {Wallraff}, {Wandkowsky}, {Watson}, {Weaver}, {Weindl}, {Weiss}, {Weldert}, {Wendt}, {Werthebach}, {Whelan}, {Whitehorn}, {Wiebe}, {Wiebusch}, {Wille}, {Williams}, {Wills}, {Wolf}, {Wood}, {Wood}, {Woschnagg}, {Wrede}, {Xu}, {Xu}, {Xu}, {Yanez}, {Yodh}, {Yoshida},
  {Yuan}, \& {Z{\"o}cklein}}]{IC2020tenyears}
{Aartsen}, M.~G., {Ackermann}, M., {Adams}, J., {et~al.} 2020, \bibinfo{title}{{Time-Integrated Neutrino Source Searches with 10 Years of IceCube Data},} \prl, 124, 051103, \dodoi{10.1103/PhysRevLett.124.051103}

\bibitem[{R. {Abbasi} {et~al.}(2022){Abbasi}, {Ackermann}, {Adams}, {Aguilar}, {Ahlers}, {Ahrens}, {Alameddine}, {Alispach}, {Alves}, {Amin}, {Andeen}, {Anderson}, {Anton}, {Arg{\"u}elles}, {Ashida}, {Axani}, {Bai}, {et~al.}}]{ngc1068_2022}
{Abbasi}, R., {Ackermann}, M., {Adams}, J., {et~al.} 2022, \bibinfo{title}{{Evidence for neutrino emission from the nearby active galaxy NGC 1068},} Science, 378, 538, \dodoi{10.1126/science.abg3395}

\bibitem[{R. {Abbasi} {et~al.}(2024){Abbasi}, {Ackermann}, {Adams}, {Agarwalla}, {Aguilar}, {Ahlers}, {Alameddine}, {Amin}, {Andeen}, {Arg{\"u}elles}, {Ashida}, {Athanasiadou}, {Ausborm}, {Axani}, {Bai}, {Balagopal V.}, {Baricevic}, {Barwick}, {Bash}, {Basu}, {Bay}, {Beatty}, {Becker Tjus}, {Beise}, {Bellenghi}, {Benning}, {BenZvi}, {Berley}, {Bernardini}, {Besson}, {Blaufuss}, {Bloom}, {Blot}, {Bontempo}, {Book Motzkin}, {Boscolo Meneguolo}, {B{\"o}ser}, {Botner}, {B{\"o}ttcher}, {Braun}, {Brinson}, {Brostean-Kaiser}, {Brusa}, {Burley}, {Butterfield}, {Campana}, {Caracas}, {Carloni}, {Carpio}, {Chattopadhyay}, {Chau}, {Chen}, {Chirkin}, {Choi}, {Clark}, {Coleman}, {Collin}, {Connolly}, {Conrad}, {Coppin}, {Corley}, {Correa}, {Cowen}, {Dave}, {De Clercq}, {DeLaunay}, {Delgado}, {Deng}, {Desai}, {Desiati}, {de Vries}, {de Wasseige}, {DeYoung}, {Diaz}, {D{\'\i}az-V{\'e}lez}, {Dierichs}, {Dittmer}, {Domi}, {Draper}, {Dujmovic}, {Dutta}, {DuVernois}, {Ehrhardt}, {Eidenschink}, {Eimer}, {Eller}, {Ellinger}, {El
  Mentawi}, {Els{\"a}sser}, {Engel}, {Erpenbeck}, {Evans}, {Evenson}, {Fan}, {Fang}, {Farrag}, {Fazely}, {Fedynitch}, {Feigl}, {Fiedlschuster}, {Finley}, {Fischer}, {Fox}, {Franckowiak}, {Fukami}, {F{\"u}rst}, {Gallagher}, {Ganster}, {Garcia}, {Garcia}, {Garg}, {Genton}, {Gerhardt}, {Ghadimi}, {Girard-Carillo}, {Glaser}, {Glauch}, {Gl{\"u}senkamp}, {Gonzalez}, {Goswami}, {Granados}, {Grant}, {Gray}, {Gries}, {Griffin}, {Griswold}, {Groth}, {G{\"u}nther}, {Gutjahr}, {Ha}, {Haack}, {Hallgren}, {Halve}, {Halzen}, {Hamdaoui}, {Minh}, {Handt}, {Hanson}, {Hardin}, {Harnisch}, {Hatch}, {Haungs}, {H{\"a}u{\ss}ler}, {Helbing}, {Hellrung}, {Hermannsgabner}, {Heuermann}, {Heyer}, {Hickford}, {Hidvegi}, {Hill}, {Hill}, {Hoffman}, {Hori}, {Hoshina}, {Hostert}, {Hou}, {Huber}, {Hultqvist}, {H{\"u}nnefeld}, {Hussain}, {Hymon}, {Ishihara}, {Iwakiri}, {Jacquart}, {Janik}, {Jansson}, {Japaridze}, {Jeong}, {Jin}, {Jones}, {Kamp}, {Kang}, {Kang}, {Kang}, {Kappes}, {Kappesser}, {Kardum}, {Karg}, {Karl}, {Karle}, {Katil}, {Katz},
  {Kauer}, {Kelley}, {Khanal}, {Khatee Zathul}, {Kheirandish}, {Kiryluk}, {Klein}, {Kochocki}, {Koirala}, {Kolanoski}, {Kontrimas}, {K{\"o}pke}, {Kopper}, {Koskinen}, {Koundal}, {Kovacevich}, \& {Kowalski}}]{Abassi2024}
{Abbasi}, R., {Ackermann}, M., {Adams}, J., {et~al.} 2024, \bibinfo{title}{{IceCube Search for Neutrino Emission from X-ray Bright Seyfert Galaxies},} arXiv e-prints, arXiv:2406.07601, \dodoi{10.48550/arXiv.2406.07601}

\bibitem[{S. {Abdollahi} {et~al.}(2023){Abdollahi}, {Ajello}, {Baldini}, {Ballet}, {Bastieri}, {Becerra Gonzalez}, {Bellazzini}, {Berretta}, {Bissaldi}, {Bonino}, {Brill}, {Bruel}, {Burns}, {Buson}, {Cameron}, {Caputo}, {Caraveo}, {Cibrario}, {Ciprini}, {Cristarella Orestano}, {Crnogorcevic}, {Cutini}, {D'Ammando}, {De Gaetano}, {Digel}, {Di Lalla}, {Di Venere}, {Dom{\'\i}nguez}, {Ramazani}, {Fegan}, {Ferrara}, {Fiori}, {Fleischhack}, {Franckowiak}, {Fukazawa}, {Fusco}, {Gammaldi}, {Gargano}, {Garrappa}, {Gasbarra}, {Gasparrini}, {Giglietto}, {Giordano}, {Giroletti}, {Green}, {Grenier}, {Guiriec}, {Gustafsson}, {Hays}, {Horan}, {Hou}, {J{\'o}hannesson}, {Kerr}, {Kocevski}, {Kuss}, {Latronico}, {Li}, {Liodakis}, {Longo}, {Loparco}, {Lorusso}, {Lott}, {Lovellette}, {Lubrano}, {Maldera}, {Manfreda}, {Mart{\'\i}-Devesa}, {Mazziotta}, {Mereu}, {Meyer}, {Michelson}, {Mizuno}, {Monzani}, {Morselli}, {Moskalenko}, {Negro}, {Omodei}, {Orlando}, {Ormes}, {Paneque}, {Panzarini}, {Perkins}, {Persic}, {Pesce-Rollins},
  {Pillera}, {Porter}, {Principe}, {Racusin}, {Rain{\`o}}, {Rando}, {Rani}, {Razzano}, {Razzaque}, {Reimer}, {Reimer}, {S{\'a}nchez-Conde}, {Parkinson}, {Scargle}, {Scotton}, {Serini}, {Sgr{\`o}}, {Siskind}, {Spandre}, {Spinelli}, {Suson}, {Tajima}, {Thompson}, {Torres}, {Valverde}, {Venters}, {Wadiasingh}, {Wagner}, \& {Wood}}]{lcr}
{Abdollahi}, S., {Ajello}, M., {Baldini}, L., {et~al.} 2023, \bibinfo{title}{{The Fermi-LAT Lightcurve Repository},} \apjs, 265, 31, \dodoi{10.3847/1538-4365/acbb6a}

\bibitem[{M. {Ackermann} {et~al.}(2015){Ackermann}, {Ajello}, {Atwood}, {Baldini}, {Ballet}, {Barbiellini}, {Bastieri}, {Becerra Gonzalez}, {Bellazzini}, {Bissaldi}, {Blandford}, {Bloom}, {Bonino}, {Bottacini}, {Brandt}, {Bregeon}, {Britto}, {Bruel}, {Buehler}, {Buson}, {Caliandro}, {Cameron}, {Caragiulo}, {Caraveo}, {Carpenter}, {Casandjian}, {Cavazzuti}, {Cecchi}, {Charles}, {Chekhtman}, {Cheung}, {Chiang}, {Chiaro}, {Ciprini}, {Claus}, {Cohen-Tanugi}, {Cominsky}, {Conrad}, {Cutini}, {D'Abrusco}, {D'Ammando}, {de Angelis}, {Desiante}, {Digel}, {Di Venere}, {Drell}, {Favuzzi}, {Fegan}, {Ferrara}, {Finke}, {Focke}, {Franckowiak}, {Fuhrmann}, {Fukazawa}, {Furniss}, {Fusco}, {Gargano}, {Gasparrini}, {Giglietto}, {Giommi}, {Giordano}, {Giroletti}, {Glanzman}, {Godfrey}, {Grenier}, {Grove}, {Guiriec}, {Hewitt}, {Hill}, {Horan}, {Itoh}, {J{\'o}hannesson}, {Johnson}, {Johnson}, {Kataoka}, {Kawano}, {Krauss}, {Kuss}, {La Mura}, {Larsson}, {Latronico}, {Leto}, {Li}, {Li}, {Longo}, {Loparco}, {Lott}, {Lovellette},
  {Lubrano}, {Madejski}, {Mayer}, {Mazziotta}, {McEnery}, {Michelson}, {Mizuno}, {Moiseev}, {Monzani}, {Morselli}, {Moskalenko}, {Murgia}, {Nuss}, {Ohno}, {Ohsugi}, {Ojha}, {Omodei}, {Orienti}, {Orlando}, {Paggi}, {Paneque}, {Perkins}, {Pesce-Rollins}, {Piron}, {Pivato}, {Porter}, {Rain{\`o}}, {Rando}, {Razzano}, {Razzaque}, {Reimer}, {Reimer}, {Romani}, {Salvetti}, {Schaal}, {Schinzel}, {Schulz}, {Sgr{\`o}}, {Siskind}, {Sokolovsky}, {Spada}, {Spandre}, {Spinelli}, {Stawarz}, {Suson}, {Takahashi}, {Takahashi}, {Tanaka}, {Thayer}, {Thayer}, {Tibaldo}, {Torres}, {Torresi}, {Tosti}, {Troja}, {Uchiyama}, {Vianello}, {Winer}, {Wood}, \& {Zimmer}}]{2015ApJ...810...14A}
{Ackermann}, M., {Ajello}, M., {Atwood}, W.~B., {et~al.} 2015, \bibinfo{title}{{The Third Catalog of Active Galactic Nuclei Detected by the Fermi Large Area Telescope},} \apj, 810, 14, \dodoi{10.1088/0004-637X/810/1/14}

\bibitem[{M.~L. {Ahnen} {et~al.}(2015){Ahnen}, {Ansoldi}, {Antonelli}, {Antoranz}, {Babic}, {Banerjee}, {Bangale}, {Barres de Almeida}, {Barrio}, {Bednarek}, {Bernardini}, {Biasuzzi}, {Biland}, {Blanch}, {Bonnefoy}, {Bonnoli}, {Borracci}, {Bretz}, {Carmona}, {Carosi}, {Chatterjee}, {Clavero}, {Colin}, {Colombo}, {Contreras}, {Cortina}, {Covino}, {Da Vela}, {Dazzi}, {De Angelis}, {De Lotto}, {de O{\~n}a Wilhelmi}, {Delgado Mendez}, {Di Pierro}, {Dominis Prester}, {Dorner}, {Doro}, {Einecke}, {Eisenacher Glawion}, {Elsaesser}, {Fern{\'a}ndez-Barral}, {Fidalgo}, {Fonseca}, {Font}, {Frantzen}, {Fruck}, {Galindo}, {Garc{\'\i}a L{\'o}pez}, {Garczarczyk}, {Garrido Terrats}, {Gaug}, {Giammaria}, {Godinovi{\'c}}, {Gonz{\'a}lez Mu{\~n}oz}, {Guberman}, {Hahn}, {Hanabata}, {Hayashida}, {Herrera}, {Hose}, {Hrupec}, {Hughes}, {Idec}, {Kodani}, {Konno}, {Kubo}, {Kushida}, {La Barbera}, {Lelas}, {Lindfors}, {Lombardi}, {L{\'o}pez}, {L{\'o}pez-Coto}, {L{\'o}pez-Oramas}, {Lorenz}, {Majumdar}, {Makariev}, {Mallot}, {Maneva},
  {Manganaro}, {Mannheim}, {Maraschi}, {Marcote}, {Mariotti}, {Mart{\'\i}nez}, {Mazin}, {Menzel}, {Miranda}, {Mirzoyan}, {Moralejo}, {Moretti}, {Nakajima}, {Neustroev}, {Niedzwiecki}, {Nievas Rosillo}, {Nilsson}, {Nishijima}, {Noda}, {Orito}, {Overkemping}, {Paiano}, {Palacio}, {Palatiello}, {Paneque}, {Paoletti}, {Paredes}, {Paredes-Fortuny}, {Persic}, {Poutanen}, {Prada Moroni}, {Prandini}, {Puljak}, {Rhode}, {Rib{\'o}}, {Rico}, {Rodriguez Garcia}, {Saito}, {Satalecka}, {Schultz}, {Schweizer}, {Shore}, {Sillanp{\"a}{\"a}}, {Sitarek}, {Snidaric}, {Sobczynska}, {Stamerra}, {Steinbring}, {Strzys}, {Takalo}, {Takami}, {Tavecchio}, {Temnikov}, {Terzi{\'c}}, {Tescaro}, {Teshima}, {Thaele}, {Torres}, {Toyama}, {Treves}, {Verguilov}, {Vovk}, {Ward}, {Will}, {Wu}, {Zanin}, {MAGIC Collaboration}, {Ajello}, {Baldini}, {Barbiellini}, {Bastieri}, {Becerra Gonz{\'a}lez}, {Bellazzini}, {Bissaldi}, {Blandford}, {Bonino}, {Bregeon}, {Bruel}, {Buson}, {Caliandro}, {Cameron}, {Caragiulo}, {Caraveo}, {Cavazzuti}, {Chiang},
  {Chiaro}, {Ciprini}, {D'Ammando}, {de Palma}, {Desiante}, {Di Venere}, {Dom{\'\i}nguez}, {Fusco}, {Gargano}, {Gasparrini}, {Giglietto}, {Giordano}, {Giroletti}, {Grenier}, {Guiriec}, {Hays}, {Hewitt}, {Jogler}, {Kuss}, {Larsson}, {Li}, {Li}, {Longo}, {Loparco}, {Lovellette}, {Lubrano}, {Maldera}, {Mayer}, {Mazziotta}, {McEnery}, {Mirabal}, {Mizuno}, {Monzani}, {Morselli}, {Moskalenko}, \& {Nuss}}]{2015ApJ...815L..23A}
{Ahnen}, M.~L., {Ansoldi}, S., {Antonelli}, L.~A., {et~al.} 2015, \bibinfo{title}{{Very High Energy {\ensuremath{\gamma}}-Rays from the Universe's Middle Age: Detection of the z = 0.940 Blazar PKS 1441+25 with MAGIC},} \apjl, 815, L23, \dodoi{10.1088/2041-8205/815/2/L23}

\bibitem[{M.~G. {Akritas} \& J. {Siebert}(1996)}]{AkritasSiebert1996}
{Akritas}, M.~G., \& {Siebert}, J. 1996, \bibinfo{title}{{A test for partial correlation with censored astronomical data},} \mnras, 278, 919, \dodoi{10.1093/mnras/278.4.919}


\bibitem[{M.~L. {Allen} {et~al.}(2024){Allen}, {Biermann}, {Chieffi}, {Frekers}, {Gergely}, {Harms}, {Jaroschewski}, {Joshi}, {Kronberg}, {Kun}, {Meli}, {Seo}, \& {Stanev}}]{2024APh...16102976A}
{Allen}, M.~L., {Biermann}, P.~L., {Chieffi}, A., {et~al.} 2024, \bibinfo{title}{{Loaded layer-cake model for cosmic ray interaction around exploding super-giant stars making black holes},} Astroparticle Physics, 161, 102976, \dodoi{10.1016/j.astropartphys.2024.102976}

\bibitem[{K.~A. {Arnaud}(1996){Arnaud}}]{Arnaud1996}
{Arnaud}, K.~A. 1996, \bibinfo{title}{{XSPEC: The First Ten Years},} in Astronomical Society of the Pacific Conference Series, Vol. 101, Astronomical Data Analysis Software and Systems V, ed. G.~H. {Jacoby} \& J.~{Barnes}, 17

\bibitem[{ {Astropy Collaboration} {et~al.}(2013){Astropy Collaboration}, {Robitaille}, {Tollerud}, {Greenfield}, {Droettboom}, {Bray}, {Aldcroft}, {Davis}, {Ginsburg}, {Price-Whelan}, {Kerzendorf}, {Conley}, {Crighton}, {Barbary}, {Muna}, {Ferguson}, {Grollier}, {Parikh}, {Nair}, {Unther}, {Deil}, {Woillez}, {Conseil}, {Kramer}, {Turner}, {Singer}, {Fox}, {Weaver}, {Zabalza}, {Edwards}, {Azalee Bostroem}, {Burke}, {Casey}, {Crawford}, {Dencheva}, {Ely}, {Jenness}, {Labrie}, {Lim}, {Pierfederici}, {Pontzen}, {Ptak}, {Refsdal}, {Servillat}, \& {Streicher}}]{2013A&A...558A..33A}
{Astropy Collaboration}, {Robitaille}, T.~P., {Tollerud}, E.~J., {et~al.} 2013, \bibinfo{title}{{Astropy: A community Python package for astronomy},} \aap, 558, A33, \dodoi{10.1051/0004-6361/201322068}

\bibitem[{ {Astropy Collaboration} {et~al.}(2018){Astropy Collaboration}, {Price-Whelan}, {Sip{\H{o}}cz}, {G{\"u}nther}, {Lim}, {Crawford}, {Conseil}, {Shupe}, {Craig}, {Dencheva}, {Ginsburg}, {VanderPlas}, {Bradley}, {P{\'e}rez-Su{\'a}rez}, {de Val-Borro}, {Aldcroft}, {Cruz}, {Robitaille}, {Tollerud}, {Ardelean}, {Babej}, {Bach}, {Bachetti}, {Bakanov}, {Bamford}, {Barentsen}, {Barmby}, {Baumbach}, {Berry}, {Biscani}, {Boquien}, {Bostroem}, {Bouma}, {Brammer}, {Bray}, {Breytenbach}, {Buddelmeijer}, {Burke}, {Calderone}, {Cano Rodr{\'\i}guez}, {Cara}, {Cardoso}, {Cheedella}, {Copin}, {Corrales}, {Crichton}, {D'Avella}, {Deil}, {Depagne}, {Dietrich}, {Donath}, {Droettboom}, {Earl}, {Erben}, {Fabbro}, {Ferreira}, {Finethy}, {Fox}, {Garrison}, {Gibbons}, {Goldstein}, {Gommers}, {Greco}, {Greenfield}, {Groener}, {Grollier}, {Hagen}, {Hirst}, {Homeier}, {Horton}, {Hosseinzadeh}, {Hu}, {Hunkeler}, {Ivezi{\'c}}, {Jain}, {Jenness}, {Kanarek}, {Kendrew}, {Kern}, {Kerzendorf}, {Khvalko}, {King}, {Kirkby}, {Kulkarni},
  {Kumar}, {Lee}, {Lenz}, {Littlefair}, {Ma}, {Macleod}, {Mastropietro}, {McCully}, {Montagnac}, {Morris}, {Mueller}, {Mumford}, {Muna}, {Murphy}, {Nelson}, {Nguyen}, {Ninan}, {N{\"o}the}, {Ogaz}, {Oh}, {Parejko}, {Parley}, {Pascual}, {Patil}, {Patil}, {Plunkett}, {Prochaska}, {Rastogi}, {Reddy Janga}, {Sabater}, {Sakurikar}, {Seifert}, {Sherbert}, {Sherwood-Taylor}, {Shih}, {Sick}, {Silbiger}, {Singanamalla}, {Singer}, {Sladen}, {Sooley}, {Sornarajah}, {Streicher}, {Teuben}, {Thomas}, {Tremblay}, {Turner}, {Terr{\'o}n}, {van Kerkwijk}, {de la Vega}, {Watkins}, {Weaver}, {Whitmore}, {Woillez}, {Zabalza}, \& {Astropy Contributors}}]{2018AJ....156..123A}
{Astropy Collaboration}, {Price-Whelan}, A.~M., {Sip{\H{o}}cz}, B.~M., {et~al.} 2018, \bibinfo{title}{{The Astropy Project: Building an Open-science Project and Status of the v2.0 Core Package},} \aj, 156, 123, \dodoi{10.3847/1538-3881/aabc4f}

\bibitem[{ {Astropy Collaboration} {et~al.}(2022){Astropy Collaboration}, {Price-Whelan}, {Lim}, {Earl}, {Starkman}, {Bradley}, {Shupe}, {Patil}, {Corrales}, {Brasseur}, {N{\"o}the}, {Donath}, {Tollerud}, {Morris}, {Ginsburg}, {Vaher}, {Weaver}, {Tocknell}, {Jamieson}, {van Kerkwijk}, {Robitaille}, {Merry}, {Bachetti}, {G{\"u}nther}, {Aldcroft}, {Alvarado-Montes}, {Archibald}, {B{\'o}di}, {Bapat}, {Barentsen}, {Baz{\'a}n}, {Biswas}, {Boquien}, {Burke}, {Cara}, {Cara}, {Conroy}, {Conseil}, {Craig}, {Cross}, {Cruz}, {D'Eugenio}, {Dencheva}, {Devillepoix}, {Dietrich}, {Eigenbrot}, {Erben}, {Ferreira}, {Foreman-Mackey}, {Fox}, {Freij}, {Garg}, {Geda}, {Glattly}, {Gondhalekar}, {Gordon}, {Grant}, {Greenfield}, {Groener}, {Guest}, {Gurovich}, {Handberg}, {Hart}, {Hatfield-Dodds}, {Homeier}, {Hosseinzadeh}, {Jenness}, {Jones}, {Joseph}, {Kalmbach}, {Karamehmetoglu}, {Ka{\l}uszy{\'n}ski}, {Kelley}, {Kern}, {Kerzendorf}, {Koch}, {Kulumani}, {Lee}, {Ly}, {Ma}, {MacBride}, {Maljaars}, {Muna}, {Murphy}, {Norman},
  {O'Steen}, {Oman}, {Pacifici}, {Pascual}, {Pascual-Granado}, {Patil}, {Perren}, {Pickering}, {Rastogi}, {Roulston}, {Ryan}, {Rykoff}, {Sabater}, {Sakurikar}, {Salgado}, {Sanghi}, {Saunders}, {Savchenko}, {Schwardt}, {Seifert-Eckert}, {Shih}, {Jain}, {Shukla}, {Sick}, {Simpson}, {Singanamalla}, {Singer}, {Singhal}, {Sinha}, {Sip{\H{o}}cz}, {Spitler}, {Stansby}, {Streicher}, {{\v{S}}umak}, {Swinbank}, {Taranu}, {Tewary}, {Tremblay}, {de Val-Borro}, {Van Kooten}, {Vasovi{\'c}}, {Verma}, {de Miranda Cardoso}, {Williams}, {Wilson}, {Winkel}, {Wood-Vasey}, {Xue}, {Yoachim}, {Zhang}, {Zonca}, \& {Astropy Project Contributors}}]{2022ApJ...935..167A}
{Astropy Collaboration}, {Price-Whelan}, A.~M., {Lim}, P.~L., {et~al.} 2022, \bibinfo{title}{{The Astropy Project: Sustaining and Growing a Community-oriented Open-source Project and the Latest Major Release (v5.0) of the Core Package},} \apj, 935, 167, \dodoi{10.3847/1538-4357/ac7c74}

\bibitem[{A.~J. {Barger} {et~al.}(2001){Barger}, {Cowie}, {Mushotzky}, \& {Richards}}]{Barger2001}
{Barger}, A.~J., {Cowie}, L.~L., {Mushotzky}, R.~F., \& {Richards}, E.~A. 2001, \bibinfo{title}{{The Nature of the Hard X-Ray Background Sources: Optical, Near-Infrared, Submillimeter, and Radio Properties},} \aj, 121, 662, \dodoi{10.1086/318742}

\bibitem[{M.~C. {Begelman} {et~al.}(1984){Begelman}, {Blandford}, \& {Rees}}]{1984RvMP...56..255B}
{Begelman}, M.~C., {Blandford}, R.~D., \& {Rees}, M.~J. 1984, \bibinfo{title}{{Theory of extragalactic radio sources},} Reviews of Modern Physics, 56, 255, \dodoi{10.1103/RevModPhys.56.255}

\bibitem[{C.~D. {Dermer} {et~al.}(2014){Dermer}, {Murase}, \& {Inoue}}]{2014JHEAp...3...29D}
{Dermer}, C.~D., {Murase}, K., \& {Inoue}, Y. 2014, \bibinfo{title}{{Photopion production in black-hole jets and flat-spectrum radio quasars as PeV neutrino sources},} Journal of High Energy Astrophysics, 3, 29, \dodoi{10.1016/j.jheap.2014.09.001}

\bibitem[{B. {Eichmann} {et~al.}(2022){Eichmann}, {Oikonomou}, {Salvatore}, {Dettmar}, \& {Becker Tjus}}]{2022ApJ...939...43E}
{Eichmann}, B., {Oikonomou}, F., {Salvatore}, S., {Dettmar}, R.-J., \& {Becker Tjus}, J. 2022, \bibinfo{title}{{Solving the Multimessenger Puzzle of the AGN-starburst Composite Galaxy NGC 1068},} \apj, 939, 43, \dodoi{10.3847/1538-4357/ac9588}

\bibitem[{F. {Haardt} \& L. {Maraschi}(1991){Haardt} \& {Maraschi}}]{1991ApJ...380L..51H}
{Haardt}, F., \& {Maraschi}, L. 1991, \bibinfo{title}{{A Two-Phase Model for the X-Ray Emission from Seyfert Galaxies},} \apjl, 380, L51, \dodoi{10.1086/186171}

\bibitem[{F. {Halzen}(2022){Halzen}}]{2022IJMPD..3130003H}
{Halzen}, F. 2022, \bibinfo{title}{{The observation of high-energy neutrinos from the cosmos: Lessons learned for multimessenger astronomy},} International Journal of Modern Physics D, 31, 2230003, \dodoi{10.1142/S0218271822300038}

\bibitem[{P.~C. {Hewett} \& V. {Wild}(2010){Hewett} \& {Wild}}]{2010MNRAS.405.2302H}
{Hewett}, P.~C., \& {Wild}, V. 2010, \bibinfo{title}{{Improved redshifts for SDSS quasar spectra},} \mnras, 405, 2302, \dodoi{10.1111/j.1365-2966.2010.16648.x}

\bibitem[{ {HI4PI Collaboration} {et~al.}(2016){HI4PI Collaboration}, {Ben Bekhti}, {Fl{\"o}er}, {Keller}, {Kerp}, {Lenz}, {Winkel}, {Bailin}, {Calabretta}, {Dedes}, {Ford}, {Gibson}, {Haud}, {Janowiecki}, {Kalberla}, {Lockman}, {McClure-Griffiths}, {Murphy}, {Nakanishi}, {Pisano}, \& {Staveley-Smith}}]{HI4PI}
{HI4PI Collaboration}, {Ben Bekhti}, N., {Fl{\"o}er}, L., {et~al.} 2016, \bibinfo{title}{{HI4PI: A full-sky H I survey based on EBHIS and GASS},} \aap, 594, A116, \dodoi{10.1051/0004-6361/201629178}

\bibitem[{D.~W. {Hogg} {et~al.}(2010){Hogg}, {Bovy}, \& {Lang}}]{Hogg2010}
{Hogg}, D.~W., {Bovy}, J., \& {Lang}, D. 2010, \bibinfo{title}{{Data analysis recipes: Fitting a model to data},} arXiv e-prints, arXiv:1008.4686, \dodoi{10.48550/arXiv.1008.4686}

\bibitem[{ {IceCube Collaboration}(2022){IceCube Collaboration}}]{ngc1068_2022_dataset}
{IceCube Collaboration}. 2022, \bibinfo{title}{{Evidence for neutrino emission from the nearby active galaxy NGC 1068},} Dataset, \dodoi{10.21234/03fq-rh11}

\bibitem[{Y. {Inoue} \& A. {Doi}(2018){Inoue} \& {Doi}}]{2018ApJ...869..114I}
{Inoue}, Y., \& {Doi}, A. 2018, \bibinfo{title}{{Detection of Coronal Magnetic Activity in Nearby Active Supermassive Black Holes},} \apj, 869, 114, \dodoi{10.3847/1538-4357/aaeb95}

\bibitem[{Y. {Inoue} {et~al.}(2020){Inoue}, {Khangulyan}, \& {Doi}}]{2020ApJ...891L..33I}
{Inoue}, Y., {Khangulyan}, D., \& {Doi}, A. 2020, \bibinfo{title}{{On the Origin of High-energy Neutrinos from NGC 1068: The Role of Nonthermal Coronal Activity},} \apjl, 891, L33, \dodoi{10.3847/2041-8213/ab7661}

\bibitem[{A. {Khatee Zathul} {et~al.}(2025){Khatee Zathul}, {Moulai}, {Fang}, & {Halzen}}]{Arifa2025}
{Khatee Zathul}, A., {Moulai}, M., {Fang}, K., \& {Halzen}, F. 2025, \bibinfo{title}{{An NGC 1068-informed Understanding of Neutrino Emission of the Active Galactic Nucleus TXS 0506+056},} \apj, 984, 54, \dodoi{10.3847/1538-4357/adc44d}

\bibitem[{T. {Kawamuro} {et~al.}(2022){Kawamuro}, {Ricci}, {Imanishi}, {Mushotzky}, {Izumi}, {Ricci}, {Bauer}, {Koss}, {Trakhtenbrot}, {Ichikawa}, {Rojas}, {Smith}, {Shimizu}, {Oh}, {den Brok}, {Baba}, {Balokovi{\'c}}, {Chang}, {Kakkad}, {Pfeifle}, {Privon}, {Temple}, {Ueda}, {Harrison}, {Powell}, {Stern}, {Urry}, \& {Sanders}}]{2022ApJ...938...87K}
{Kawamuro}, T., {Ricci}, C., {Imanishi}, M., {et~al.} 2022, \bibinfo{title}{{BASS XXXII: Studying the Nuclear Millimeter-wave Continuum Emission of AGNs with ALMA at Scales {\ensuremath{\lesssim}}100-200 pc},} \apj, 938, 87, \dodoi{10.3847/1538-4357/ac8794}

\bibitem[{K.~I. {Kellermann} {et~al.}(2007){Kellermann}, {Kovalev}, {Lister}, {Homan}, {Kadler}, {Cohen}, {Ros}, {Zensus}, {Vermeulen}, {Aller}, \& {Aller}}]{2007Ap&SS.311..231K}
{Kellermann}, K.~I., {Kovalev}, Y.~Y., {Lister}, M.~L., {et~al.} 2007, \bibinfo{title}{{Doppler boosting, superluminal motion, and the kinematics of AGN jets},} \apss, 311, 231, \dodoi{10.1007/s10509-007-9622-5}

\bibitem[{B.~C. {Kelly}(2007){Kelly}}]{Kelly2007}
{Kelly}, B.~C. 2007, \bibinfo{title}{{Some Aspects of Measurement Error in Linear Regression of Astronomical Data},} \apj, 665, 1489, \dodoi{10.1086/519947}

\bibitem[{M.~J. {Koss} {et~al.}(2022){Koss}, {Ricci}, {Trakhtenbrot}, {Oh}, {den Brok}, {Mej{\'\i}a-Restrepo}, {Stern}, {Privon}, {Treister}, {Powell}, {Mushotzky}, {Bauer}, {Ananna}, {Balokovi{\'c}}, {B{\"a}r}, {Becker}, {Bessiere}, {Burtscher}, {Caglar}, {Congiu}, {Evans}, {Harrison}, {Heida}, {Ichikawa}, {Kamraj}, {Lamperti}, {Pacucci}, {Ricci}, {Riffel}, {Rojas}, {Schawinski}, {Temple}, {Urry}, {Veilleux}, \& {Williams}}]{2022ApJS..261....2K}
{Koss}, M.~J., {Ricci}, C., {Trakhtenbrot}, B., {et~al.} 2022, \bibinfo{title}{{BASS. XXII. The BASS DR2 AGN Catalog and Data},} \apjs, 261, 2, \dodoi{10.3847/1538-4365/ac6c05}

\bibitem[{Y.~Y. {Kovalev} {et~al.}(2025){Kovalev}, {Pushkarev}, {G{\'o}mez}, {Homan}, {Lister}, {Livingston}, {Pashchenko}, {Plavin}, {Savolainen}, \& {Troitsky}}]{Kovalev2025}
{Kovalev}, Y.~Y., {Pushkarev}, A.~B., {G{\'o}mez}, J.~L., {et~al.} 2025, \bibinfo{title}{{Looking into the jet cone of the neutrino-associated very high-energy blazar PKS 1424+240},} \aap, 700, L12, \dodoi{10.1051/0004-6361/202555400}

\bibitem[{E. {Kun} {et~al.}(2023){Kun}, {Bartos}, {Becker Tjus}, {Biermann}, {Franckowiak}, {Halzen}, \& {Mez{\H{o}}}}]{Kun2023}
{Kun}, E., {Bartos}, I., {Becker Tjus}, J., {et~al.} 2023, \bibinfo{title}{{Searching for temporary gamma-ray dark blazars associated with IceCube neutrinos},} \aap, 679, A46, \dodoi{10.1051/0004-6361/202346710}

\bibitem[{E. {Kun} {et~al.}(2021){Kun}, {Bartos}, {Becker Tjus}, {Biermann}, {Halzen}, \& {Mez{\H{o}}}}]{2021ApJ...911L..18K}
{Kun}, E., {Bartos}, I., {Becker Tjus}, J., {et~al.} 2021, \bibinfo{title}{{Cosmic Neutrinos from Temporarily Gamma-suppressed Blazars},} \apjl, 911, L18, \dodoi{10.3847/2041-8213/abf1ec}

\bibitem[{E. {Kun} {et~al.}(2024){Kun}, {Bartos}, {Tjus}, {Biermann}, {Franckowiak}, {Halzen}, {del Palacio}, \& {Woo}}]{Kun2024}
{Kun}, E., {Bartos}, I., {Tjus}, J.~B., {et~al.} 2024, \bibinfo{title}{{Possible correlation between unabsorbed hard x rays and neutrinos in radio-loud and radio-quiet active galactic nuclei},} \prd, 110, 123014, \dodoi{10.1103/PhysRevD.110.123014}


\bibitem[{J.-J. Luo {et~al.}(2026)Luo, Lu, \& Liang}]{LuoLuLiang2026}
Luo, J.-J., Lu, M.-X., \& Liang, Y.-F. 2026, On the Apparent Correlation between X-ray and Neutrino Luminosities of Active Galactic Nuclei, \doarXiv{2605.13588}

\bibitem[{J.~S. {Miller} {et~al.}(1978){Miller}, {French}, \& {Hawley}}]{1978bllo.conf..176M}
{Miller}, J.~S., {French}, H.~B., \& {Hawley}, S.~A. 1978, \bibinfo{title}{{Optical spectra of BL Lacertae objects.},} in BL Lac Objects, ed. A.~M. {Wolfe}, 176--187

\bibitem[{K. {Murase} {et~al.}(2020){Murase}, {Kimura}, \& {M{\'e}sz{\'a}ros}}]{2020PhRvL.125a1101M}
{Murase}, K., {Kimura}, S.~S., \& {M{\'e}sz{\'a}ros}, P. 2020, \bibinfo{title}{{Hidden Cores of Active Galactic Nuclei as the Origin of Medium-Energy Neutrinos: Critical Tests with the MeV Gamma-Ray Connection},} \prl, 125, 011101, \dodoi{10.1103/PhysRevLett.125.011101}

\bibitem[{A. {Neronov} {et~al.}(2024){Neronov}, {Savchenko}, \& {Semikoz}}]{Neronov2024}
{Neronov}, A., {Savchenko}, D., \& {Semikoz}, D.~V. 2024, \bibinfo{title}{{Neutrino Signal from a Population of Seyfert Galaxies},} \prl, 132, 101002, \dodoi{10.1103/PhysRevLett.132.101002}

\bibitem[{K. {Oh} {et~al.}(2018){Oh}, {Koss}, {Markwardt}, {Schawinski}, {Baumgartner}, {Barthelmy}, {Cenko}, {Gehrels}, {Mushotzky}, {Petulante}, {Ricci}, {Lien}, \& {Trakhtenbrot}}]{2018ApJS..235....4O}
{Oh}, K., {Koss}, M., {Markwardt}, C.~B., {et~al.} 2018, \bibinfo{title}{{The 105-Month Swift-BAT All-sky Hard X-Ray Survey},} \apjs, 235, 4, \dodoi{10.3847/1538-4365/aaa7fd}

\bibitem[{P. {Padovani}(1992){Padovani}}]{Padovani1992}
{Padovani}, P. 1992, \bibinfo{title}{{A statistical analysis of complete samples of BL Lacertae objects},} \aap, 256, 399


\bibitem[{P. {Padovani} {et~al.}(2017){Padovani}, {Alexander}, {Assef}, {De Marco}, {Giommi}, {Hickox}, {Richards}, {Smol{\v{c}}i{\'c}}, {Hatziminaoglou}, {Mainieri}, \& {Salvato}}]{2017A&ARv..25....2P}
{Padovani}, P., {Alexander}, D.~M., {Assef}, R.~J., {et~al.} 2017, \bibinfo{title}{{Active galactic nuclei: what's in a name?},} \aapr, 25, 2, \dodoi{10.1007/s00159-017-0102-9}

\bibitem[{S. {Paiano} {et~al.}(2016){Paiano}, {Landoni}, {Falomo}, {Scarpa}, \& {Treves}}]{2016MNRAS.458.2836P}
{Paiano}, S., {Landoni}, M., {Falomo}, R., {Scarpa}, R., \& {Treves}, A. 2016, \bibinfo{title}{{On the redshift of the very high-energy gamma-ray BL Lac object S2 0109+22},} \mnras, 458, 2836, \dodoi{10.1093/mnras/stw472}

\bibitem[{A.~V. {Plavin} {et~al.}(2024){Plavin}, {Burenin}, {Kovalev}, {Lutovinov}, {Starobinsky}, {Troitsky}, \& {Zakharov}}]{2024JCAP...05..133P}
{Plavin}, A.~V., {Burenin}, R.~A., {Kovalev}, Y.~Y., {et~al.} 2024, \bibinfo{title}{{Hard X-ray emission from blazars associated with high-energy neutrinos},} \jcap, 2024, 133, \dodoi{10.1088/1475-7516/2024/05/133}

\bibitem[{C. {Ricci} {et~al.}(2023){Ricci}, {Chang}, {Kawamuro}, {Privon}, {Mushotzky}, {Trakhtenbrot}, {Laor}, {Koss}, {Smith}, {Gupta}, {Dimopoulos}, {Aalto}, \& {Ros}}]{2023ApJ...952L..28R}
{Ricci}, C., {Chang}, C.-S., {Kawamuro}, T., {et~al.} 2023, \bibinfo{title}{{A Tight Correlation between Millimeter and X-Ray Emission in Accreting Massive Black Holes from <100 mas Resolution ALMA Observations},} \apjl, 952, L28, \dodoi{10.3847/2041-8213/acda27}

\bibitem[{N. {Sahakyan} {et~al.}(2024){Sahakyan}, {Harutyunyan}, {Gasparyan}, \& {Israyelyan}}]{2024MNRAS.528.5990S}
{Sahakyan}, N., {Harutyunyan}, G., {Gasparyan}, S., \& {Israyelyan}, D. 2024, \bibinfo{title}{{Broad-band study of gamma-ray blazars at redshifts z = 2.0-2.5},} \mnras, 528, 5990, \dodoi{10.1093/mnras/stae273}

\bibitem[{J. {Wilms} {et~al.}(2000){Wilms}, {Allen}, \& {McCray}}]{wilm}
{Wilms}, J., {Allen}, A., \& {McCray}, R. 2000, \bibinfo{title}{{On the Absorption of X-Rays in the Interstellar Medium},} \apj, 542, 914, \dodoi{10.1086/317016}

\end{thebibliography}

\end{document}